\documentclass[twocolumn,twocolappendix]{aastex63}

\usepackage{amsmath,mathtools,mathrsfs,amssymb}
\usepackage{hyperref}
\usepackage{verbatim}
\usepackage{xspace}
\usepackage{graphicx}
\usepackage{longtable}
\usepackage[T1]{fontenc}
\graphicspath{{./}{figures/}}

\newcommand{\harm}[0]{{\tt{harm}}\xspace}
\newcommand{\iharm}[0]{{\tt{iharm3d}}\xspace}
\newcommand{\kharma}[0]{{KHARMA}\xspace}

\def\v3{\texttt{v3}}
\def\sgra{Sgr~A$^*$\xspace}
\def\m87{M87$^*$\xspace}

\defcitealias{M87PaperI}{EHTC~M87~I}
\defcitealias{M87PaperII}{EHTC~M87~II}
\defcitealias{M87PaperIII}{EHTC~M87~III}
\defcitealias{M87PaperIV}{EHTC~M87~IV}
\defcitealias{M87PaperV}{EHTC~M87~V}
\defcitealias{M87PaperVI}{EHTC~M87~VI}
\defcitealias{M87PaperVII}{EHTC~M87~VII}
\defcitealias{M87PaperVIII}{EHTC~M87~VIII}
\defcitealias{M87PaperIX}{EHTC~M87~IX}

\defcitealias{SgrAPaperI}{EHTC~SgrA~I}
\defcitealias{SgrAPaperII}{EHTC~SgrA~II}
\defcitealias{SgrAPaperIII}{EHTC~SgrA~III}
\defcitealias{SgrAPaperIV}{EHTC~SgrA~IV}
\defcitealias{SgrAPaperV}{EHTC~SgrA~V}
\defcitealias{SgrAPaperVI}{EHTC~SgrA~VI}
\defcitealias{SgrAPaperVII}{EHTC~SgrA~VII}
\defcitealias{SgrAPaperVIII}{EHTC~SgrA~VIII}

\begin{document}

\title{A Survey of General Relativistic Magnetohydrodynamic Models for Black Hole Accretion Systems}

\correspondingauthor{Vedant Dhruv}
\email{vdhruv2@illinois.edu}

\author[0000-0001-6765-877X]{Vedant Dhruv}
\affiliation{Department of Physics, University of Illinois, 1110 West Green St., Urbana, IL 61801, USA}
\affiliation{Illinois Center for Advanced Study of the Universe, 1110 West Green Street, Urbana, IL 61801, USA}

\author[0000-0002-0393-7734]{Ben Prather}
\affiliation{CCS-2, Los Alamos National Laboratory, P.O. Box 1663, Los Alamos, NM 87545, USA}

\author[0000-0001-6952-2147]{George~N.~Wong}
\affiliation{School of Natural Sciences, Institute for Advanced Study, 1 Einstein Drive, Princeton, NJ 08540, USA}
\affiliation{Princeton Gravity Initiative, Princeton University, Princeton, New Jersey 08544, USA}

\author[0000-0000-0000-0000]{Charles F. Gammie}
\affiliation{Department of Physics, University of Illinois, 1110 West Green St., Urbana, IL 61801, USA}
\affiliation{Illinois Center for Advanced Study of the Universe, 1110 West Green Street, Urbana, IL 61801, USA}
\affiliation{Department of Astronomy, University of Illinois, 1002 West Green St., Urbana, IL 61801, USA}
\begin{abstract}
    
General Relativistic Magnetohydrodynamics (GRMHD) simulations are an indispensable tool in studying accretion onto compact objects. The Event Horizon Telescope (EHT) frequently uses libraries of ideal GRMHD simulations to interpret polarimetric, event-horizon-scale observations of supermassive black holes at the centers of galaxies. In this work, we present a library of ten non-radiative, ideal GRMHD simulations that were utilized by the EHT Collaboration in their analysis of Sagittarius A*. The parameter survey explores both low (SANE) and high (MAD) magnetization states across five black hole spins $a_{*}=-15/16,-1/2,0,+1/2,+15/16$ where each simulation was run out to $30,000\hspace{0.1cm}\mathrm{GM/c}^{3}$. We find the angular momentum and energy flux in SANE simulations closely matches  the thin-disk value, with minor deviations in prograde models due to fluid forces. This leads to spin equilibrium around $a_{*}\sim0.94$, consistent with previous studies. We study the flow of conserved quantities in our simulations and find mass, angular momentum, and energy transport in SANE accretion flows to be primarily inward and fluid-dominated. MAD models produce powerful jets with outflow efficiency $>1$ for $a_{*}=+0.94$, leading to black hole spin-down in prograde cases.  We observe outward directed energy and angular momentum fluxes on the horizon, as expected for the Blandford-Znajek mechanism.  MAD accretion flows are sub-Keplerian and exhibit greater variability than their SANE counterpart. They are also hotter than SANE disks within $r\lesssim 10\hspace{0.1cm}\mathrm{GM/c}^{2}$. This study is accompanied by a public release of simulation data at \url{http://thz.astro.illinois.edu/}.  
    
\end{abstract}

\section{Introduction}

It is now widely believed that supermassive black holes reside at the centers of nearly all galaxies \citep{lynden-bell_agn_1969,kormendy_richstone_smbh_1995,richstone_smbh_1998}, are fed by nearby stars and gas clouds, and play a key role in the evolution of the host galaxy (see, e.g., \citealt{silk_rees_mass_bulge_1998,magorrian_relations_1998,king_mass_velocity_dispersion_2003,hopkins_cosmo_feddback_2024}). Of these, \sgra and \m87---the putative supermassive black holes at the centers of our Galaxy and M87 galaxy respectively---subtend the largest angle on the sky ($\sim50\mu as$). Additionally, near-horizon emission from these sources peaks at millimeter wavelengths, making them suitable for ground-based VLBI observation \citep{doeleman_eht_white_paper_2009}.

Over the past five years, the EHT---a global array of millimeter wavelength telescopes---has released horizon-scale, total intensity and polarimetric images of \m87 \citep[][hereafter EHTC~M87~I--IX]{M87PaperI,M87PaperII,M87PaperIII,M87PaperIV,M87PaperV,M87PaperVI,M87PaperVII,M87PaperVIII,M87PaperIX} and \sgra \citep[][hereafter EHTC~SgrA~I--VIII]{SgrAPaperI,SgrAPaperII,SgrAPaperIII,SgrAPaperIV,SgrAPaperV,SgrAPaperVI,SgrAPaperVII,SgrAPaperVIII}. These images show a bright ring of emission encircling a central depression in brightness, or ``shadow'', and are broadly consistent with models of geometrically thick, hot, two-temperature, advection-dominated accretion flows \citep{ichimaru_bimodal_1977,rees_hot_tori_1982,narayan_adaf_1994,narayan_adaf_1995,narayan_sgra_spectrum_adaf_1995}. 

The theoretical interpretation accompanying these observations is based on a large set of synthetic images and spectral energy distributions generated by carrying out general relativistic radiative transfer calculations on data from GRMHD simulations of supermassive black hole accretion. Summary statistics derived from these electromagnetic observables are compared against observations allowing inferences to be drawn about the source \citepalias{M87PaperV, SgrAPaperV}. The observations support magnetically dominated accretion flows and provide constraints on the black hole spin, accretion rate, and the electron-to-ion temperature ratio in the plasma. In the case of \sgra, the analysis also provides constraints on the observer inclination relative to the angular momentum of the accretion flow (\citetalias{SgrAPaperV}).

The EHT theory pipeline relies on a \textit{library} of GRMHD simulations of black hole accretion. The ``canonical'' set of simulations in these analyses solves the equations of ideal GRMHD for a single-temperature fluid \citep{gammie_harm_2003, de_villiers_grmhd_2003,del_zanna_echo_grmhd_2007,porth_cc_2019}, starting from similar initial conditions across a range of black hole spins and magnetization states (see Section \ref{subsec:initial_conditions}), using multiple codes \citep{sadowski_koral_2013,porth_bhac_2017,Prather_2021_iharm3d,liska_hamr_2022}. The \sgra and \m87 analyses also incorporates a curated selection of simulations featuring varying initial conditions, such as tilted accretion disks  \citep{chatterjee_tilted_images_2020} and stellar wind-fed models \citep{ressler_abinitio_2020}, and alternative approaches to modeling electron temperature \citep{ryan_two_temp_m87_2018,dexter_parameter_2020}.

In this paper we present a version of the GRMHD library, which we will refer to as \v3, generated by the performance-portable, GPU-enabled code \kharma that was instrumental in analyzing horizon-scale observations of the galactic center \citepalias{SgrAPaperV,SgrAPaperVIII}. We limit our analysis to the fluid-level data products, i.e., the spatio-temporal information of the fluid state and the magnetic fields prior to any radiative transfer post-processing. In Section \ref{sec:grmhd} we provide a primer on ideal GRMHD and go over some of the assumptions when modelling low-luminosity active galactic nuclei (LLAGN) using ideal GRMHD. In Section \ref{sec:numerics_simulations} we introduce \kharma, outline the initial conditions and model parameters for the \v3 library, and discuss the numerical shortcomings of our algorithm. In Section \ref{sec:discussions_results}, we discuss trends in our GRMHD simulation library by analyzing time-series and time-averaged data. In Section \ref{sec:data_products} we provide a URL for the GRMHD data used in this paper. We conclude in Section \ref{sec:summary} by summarizing our findings and listing the limitations of our model library.
\section{General relativistic magnetohydrodynamics (GRMHD)}
\label{sec:grmhd}

Since the present work focuses on GRMHD data products, we provide a brief overview of the governing equations, nomenclature of fluid variables, and discuss the assumptions that go into our black hole accretion model. We adopt a set of units where $ GM = c = 1$. For the electromagnetic sector we use Lorentz--Heaviside units, which, similar to CGS, set the vacuum permittivity and permeability to unity ($\varepsilon_0 = \mu_0 = 1$). However, unlike CGS, Lorentz--Heaviside unit system absorbs factors of $\sqrt{4\pi}$ into the definition of the electromagnetic fields.

\subsection{Equations of ideal GRMHD}

The governing equations of ideal GRMHD include (i)  conservation of particle number, (ii) conservation of energy and momentum (stress-energy tensor) and (iii) the source-free half of Maxwell's equations.  When expressed in a covariant manner these take the form, 
\begin{align}\label{eqn:ideal_mhd_equations_covariant_form}
    \nabla_{\mu}(\rho u^{\mu}) &= 0, \nonumber \\
    \nabla_{\mu}T^{\mu\nu} &= 0, \\
    \nabla_{\nu}\text{*}F^{\mu\nu} &= 0, \nonumber
\end{align}
where $\text{*}F^{\mu\nu}$ is the dual of the electromagnetic tensor $F^{\mu\nu}$, $\text{*}F^{\mu\nu} = \frac{1}{2}\epsilon^{\mu\nu\alpha\beta}F_{\alpha\beta}$. Note that Greek indices run over all four dimensions (0, 1, 2, 3) while Latin indices run over the spatial dimensions (1, 2, 3). $\rho$ is the rest-mass density of the fluid, i.e., it is the density of the fluid as measured by an observer comoving with the fluid (also known as `fluid frame'), $
u^{\mu}$ is the fluid 4-velocity and $T^{\mu\nu}$ is the ideal MHD stress-energy tensor,
\begin{equation} \label{eqn:ideal_stress_energy_tensor}
    T^{\mu\nu} = (\rho + u + p_g + b^2)u^{\mu}u^{\nu} + (p_g +\frac{b^2}{2})g^{\mu\nu} -b^{\mu}b^{\nu}.
\end{equation}
$u$ and $p_g$ are the fluid specific internal energy and pressure respectively, as measured in the fluid frame. $b^{\mu}$ is the magnetic-field 4-vector, $b^{\mu} \equiv \frac{1}{2} \epsilon^{\mu\nu\kappa\lambda}u_{\nu}F_{\lambda\kappa}$. In the fluid frame this has the more intuitive form $b^{\mu} \rightarrow (0,\boldsymbol{B})$, where $\boldsymbol{B}$ is the magnetic field 3-vector measured in the ``lab frame''. This relates to the magnetic field measured by the normal observer\footnote{The 4-velocity of the normal observer is given by $\eta_{\mu}=(-\alpha,0,0,0)$ where $\alpha\equiv1/\sqrt{-g_{tt}}$ is the lapse.} ($\boldsymbol{\mathcal{B}}$) as $B^{i}=\mathcal{B}^{i}/\alpha=\text{*}F^{it}$. In an arbitrary frame the components of $b^{\mu}$ are related to $\boldsymbol{B}$ by,
\begin{align}\label{eqn:magnetic_field_three_four_vector_relation}
\begin{split}
    b^{t} &= g_{i\mu}B^{i}u^{\mu},\\
    b^{i} &= (B^{i} + b^{t}u^{i})/u^{t}.
\end{split}
\end{align}
See Appendix B in \cite{chael_polarimetry_2023} for a more in-depth discussion on degenerate electromagnetic fields in ideal GRMHD. In the ideal MHD limit, electric fields vanish in the fluid frame. Consequently,  it can be shown that the 4-vectors $u^{\mu}$ and $b^{\mu}$ satisfy $b^{\mu}u_{\mu}=0$.

We rewrite the equations of ideal GRMHD (Equation \ref{eqn:ideal_mhd_equations_covariant_form}) in conservation form,
\begin{align}\label{eqn:ideal_mhd_equations_conservation_form}
    \partial_t\big(\sqrt{-g}\rho u^t\big) &= -\partial_i\big(\sqrt{-g}\rho u^i\big) \nonumber \\
    \partial_t\big(\sqrt{-g}T^t_{\nu}\big) &= -\partial_i\big(\sqrt{-g}T^i_{\nu}\big) + \sqrt{-g}T^{\kappa}_{\lambda}\Gamma^{\lambda}_{\nu\kappa}\\
    \partial_t\big(\sqrt{-g}B^i\big) &= -\partial_j\big(\sqrt{-g}(b^j u^i - b^i u^j)\big), \nonumber
\end{align}
and,
\begin{equation}\label{eqn:no_monopole_constraint}
    \partial_i\big(\sqrt{-g}B^i\big) = 0,
\end{equation}
where we have expressed the equations in a coordinate basis $x^{\mu}$. Equation (\ref{eqn:no_monopole_constraint}) is the divergence-free criterion for the magnetic fields. Here $g\equiv\text{det}(g_{\mu\nu})$ is the determinant of the covariant metric and $\Gamma^{\lambda}_{\mu\nu}$ is the Christoffel symbol. We assume a $\hat{\gamma}$-law equation of state (hereafter EoS), $p_g = (\hat{\gamma}-1)u$, where $\hat{\gamma}$ is the fluid adiabatic index. The other, inhomogeneous, half of Maxwell's equations,
\begin{equation}
    \nabla_{\nu}F^{\mu\nu} = J^{\mu},
\end{equation}
determines the 4-current $J^{\mu}$.

\subsection{Assumptions and caveats}\label{subsec:grmhd_asusmptions_caveats}

The Coulomb collisional mean free paths of the ions and electrons in models of RIAFs are much larger than the length scales associated with the accretion disks \citep{mahadevan_quataert_1997}. This has two significant implications on the nature of the plasma that is of interest to us.

First, low collisionality suggests that the fluid approach is not sufficient to capture the relevant physics and a kinetic approach is necessary. However,  Particle-In-Cell (PIC) simulations have shown that microscale instabilities give rise to wave-particle interactions which in turn may increase the \textit{effective} collision rate \citep{Kunz_2014, sironi_narayan_2015, sironi_2015, riquelme_pic_2015, Riquelme_2016, Riquelme_2018, inchingolo_kinetic_mri_2018, bacchini_shearing_box_2024}. This suggests that non-ideal effects such as thermal conduction and viscosity may play an important role in the dynamics and thermodynamics of the accreting plasma. \cite{chandra_extended_2015, foucart_nonideal_2017} construct one such model, where the dissipative processes are anisotropic with respect to the local magnetic field. They find the non-ideal effects to not appreciably change the time-averaged structure of the flow for the closure parameters that were considered.

Second, the large thermalization times between the ions and electrons, differences in the heating mechanisms for the two species, and efficient radiative cooling for electrons, implies a two-temperature plasma with the ions being hotter than the electrons \citep{Eardley_1975, shapiro_1976, rees_hot_tori_1982, mahadevan_quataert_1997}. Global simulations of radiatively inefficient accretion flow typically simulate a single temperature fluid and assign electron temperatures during post-processing by apportioning the fluid internal energy based on various electron heating mechanisms and some assumption about the electron distribution function \citep{moscibrodzka_general_2016}. \cite{ressler_electron_2015} formulated a sub-grid electron heating scheme where the electrons are modelled as a passive fluid that does not backreact onto the gas. A fraction of the numerical dissipation is appropriated based on the local fluid state and magnetic field \citep{howes_prescription_2010, rowan_electron_2017, werner_non-thermal_2018, kawazura_thermal_2019,zhdankin_production_2021} and is used to heat the electrons. While this electron entropy tracking procedure is implemented in our code, we do not consider it in this study due to compute and storage limitations at the time. However, a select set of such models was considered in \citetalias{SgrAPaperV} to study how a more detailed treatment of the electron thermodynamics might influence 230 GHz lightcurve variability.  

In systems of interest to us we expect the advective motion of the magnetic fields to dominate over diffusion, i.e., $\text{Re}_{\text{m}}\gg1$ ($\mathrm{Re}_{\mathrm{m}}$ is the magnetic Reynolds number) and ignore explicit resistivity. However, resistive effects are necessary if one wishes to explicitly capture magnetic reconnection that drives particles towards a non-thermal distribution and has been suggested as a possible explanation of near-infrared (NIR) and X-ray flares \citep{ripperda_reconnection_hotspots_2020,ripperda_flares_2022,nathanail_plasmoid_grmhd_flares_2020,scepi_sgra_xray_flares_2022,galishnikova_grpic_2023,mellah_magnetospheres_pic_reconnection_2023,vos_plasmoids_harris_sheets_grmhd_2024}.


For systems accreting far below the Eddington rate, $\dot{m}\equiv\dot{M}/\dot{M}_{\mathrm{Edd}}\ll1$ ($\dot{M}_{\mathrm{Edd}}=2.2\times10^{-8} (M/M_{\odot})M_{\odot}\hspace{0.05in}\text{yr}^{-1}$, where we have chosen nominal efficiency $\eta=0.1$), radiative cooling timescales are much longer than inflow timescales and radiation feedback can be ignored. For $\dot{m}\gtrsim10^{-6}$, synchrotron emission and Compton scattering become important, and at higher accretion rates $\dot{m}\sim10^{-4}$, Coulomb collisions also come into the picture \citep{Dibi_2012, Ryan_2017_radiative_efficiency}. The accretion rate at the galactic center is believed to be much below this (\citealp{Quataert_2000_mdot_saga, Bower_2003_lp_saga}, \citetalias{SgrAPaperV}) and we do not include radiative cooling.

In our models we set $\hat{\gamma}$ to a constant value across the simulation domain. This is not true in general since $\hat{\gamma}$ depends on the local conditions of the fluid. We refer interested readers to \cite{Mignone_2005, mignone_equation_2007, Choi_2010_eos, Mizuno_2013_eos, Sadowski_2017} for discussions about the limitations of the $\hat{\gamma}$-law EoS, physically motivated alternate EoSs, and a scheme to self-consistently evolve adiabatic indices of the electrons and ions.
\section{Numerics and simulations}
\label{sec:numerics_simulations}

The quasilinear system of equations that describes ideal GRMHD (Equations \ref{eqn:ideal_mhd_equations_conservation_form}), are hyperbolic and therefore well-posed \citep{lichnerowicz_1967, anile_mathematical_structure_rmhd_1987, anile_mhd_1990, komissarov_godunov-type_1999}. The \harm algorithm \citep{gammie_harm_2003} is a conservative scheme and consequently (i) ensures convergence to a weak solution of the problem in 1D, if the problem is convergent at all \citep{lax_wendroff_conservation_1960} and (ii) can capture strong shocks \citep{hou_1994}.

\subsection{Code details}

The simulation suite presented in this work was generated by \kharma (Kokkos-based High-Accuracy Relativistic Magnetohydrodynamics with Adaptive Mesh Refinement (AMR); \citealt{prather_kharma_2024})\footnote{The code is open-source and available at \url{https://github.com/AFD-Illinois/kharma}}, a C++17 rewrite of \iharm \citep{Prather_2021_iharm3d} that leverages the Kokkos framework \citep{CARTER_2014_kokkos, Trott_2021_kokkos, Trott_2022_kokkos} to run efficiently on CPUs and GPUs. It utilizes the Parthenon framework \citep{grete_parthenon_2022} to (i) achieve block-structured static and adaptive mesh refinement, (ii) introduce flexibility and modularity in the code through the use of dynamic task lists and packages, and (iii) leverage its interface for MPI communication and parallel HDF5 I/O operations.

\kharma uses a second-order predictor-corrector scheme to step forward in time. The magnetohydrodynamic fields are stored at zone centers and are reconstructed at zone faces using the WENO5 scheme of \cite{jiang1996} to compute face-centered transport fluxes. The divergence-free condition (Equation \ref{eqn:no_monopole_constraint}) is maintained to machine precision by employing the flux-interpolated constrained transport (flux-CT) scheme of \cite{toth_2000_divb} \footnote{Since the \v3 library was generated, \kharma now supports the face-centered constrained transport scheme described in \cite{stone_gardiner_godunov_mhd_2009}.}. \kharma solves the Riemann problem at zone faces using the LLF solver \citep{rusanov_1962}. \kharma evolves 8 scalar fields for ideal GRMHD; these are the set of conserved variables,
\begin{equation}\label{eqn:conserved_variables}
\boldsymbol{U} \equiv \sqrt{-g} (\rho u^t, T^t_t, T^t_i, B^i).
 \end{equation}
Additionally, it keeps the corresponding primitive fields,
\begin{equation}\label{eqn: primitive_variables}
\boldsymbol{P} \equiv (\rho, u, \Tilde{u}^i, B^i),
\end{equation}
in lockstep. $\Tilde{u}^i$ is the fluid velocity as measured by the normal observer scaled by the Lorentz factor. \footnote{In numerical relativity terminology, the fluid velocity relative to a normal (Eulerian) observer with velocity $n^{\mu}=\frac{1}{\alpha}(1,-\beta^{i})$ is given by $u'^{i} = \frac{u^{i}}{\Gamma}+\frac{\beta^{i}}{\alpha}$, where $\beta^{i}\equiv g^{ti}\alpha^2$ is the shift. Note that $\Gamma$ is the Lorentz factor of the fluid with respect to the normal observer, $\Gamma=-n_{\mu}u^{\mu}$. In \harm we instead consider $\Tilde{u}^i\equiv \Gamma u'^{i}$ for numerical stability reasons.}


\subsection{Initial conditions}\label{subsec:initial_conditions}

We initialize the fluid sector of the simulations with a Fishbone-Moncrief (FM) torus (\citealt{fishbone_relativistic_1976}) that is parameterized with the inner radius of the disk, $r_{\text{in}}$, and the radius at maximum pressure, $r_{\text{max}}$. The thermal energy of the fluid is perturbed to seed development of instabilities such as the MRI (\citealt{mri_1991}). A detailed description of the implementation of the FM torus  can be found in Appendix A of \cite{Wong_2022_patoka}.

\begin{figure}
\centering
\includegraphics[,width=\linewidth]{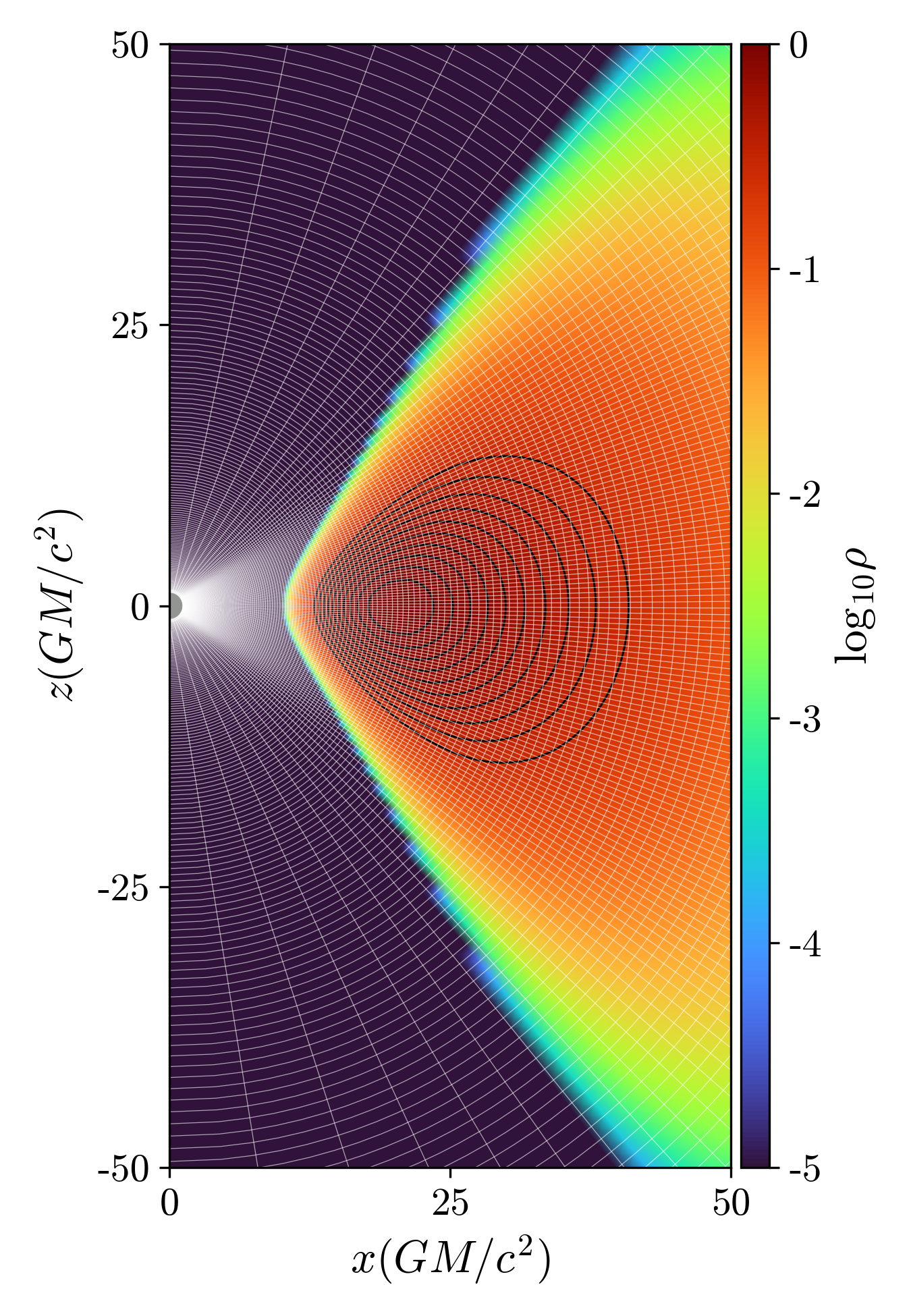}
\caption{Azimuthal/poloidal slice of initial conditions along with the grid geometry. The color scale denotes the logarithm of the rest-mass density and as an example we plot the magnetic fields structure (in black) for a SANE disk. The grid zone boundaries are represented by the white mesh. The grid zones are concentrated towards the equatorial plane and at smaller radii, where we expect most of the relevant physics to occur.}
\label{fig:grid_init}
\end{figure}

The electromagnetic sector of the simulations is initialized with a single poloidal loop of magnetic field by specifying the toroidal component of the magnetic four-vector potential $A_{\phi}$. The strength and structure of the initial field dictates the magnetic flux $\Phi_{\mathrm{BH}}$ (also expressed in terms of the dimensionless magnetic flux $\phi_{b}\sim\Phi_{\text{BH}}/\sqrt{\dot{M}}$) threading the event horizon at steady state and gives rise to two qualitatively different modes of accretion. When $\phi_{b}\sim\phi_{b,\text{crit}}$ (where $\phi_{b,\text{crit}}\sim 15$ is the critical value at which the outward magnetic pressure balances the inward fluid pressure \footnote{The value of $\phi_{b,\mathrm{crit}}$, a measure of the maximum allowable magnetic flux trapped at the horizon, is believed to depend on the black hole spin and the adiabatic index of the accreting fluid.}; \citealt{tchekhovskoy_efficient_2011}) we obtain a magnetically arrested disk (MAD) (\citealt{bisnovatyi-Kogan_1974, bisnovatyi-kogan_1976, narayan_mad_2003, igumenshchev_2003}); the initial magnetic 4-vector potential for which is,
\begin{equation}\label{eqn: mad_four_potential}
    A_{\phi} = \text{max}\Bigg[\frac{\rho}{\rho_{\text{max}}}\Big(\frac{r}{r_{\text{in}}}\text{sin}~\theta\Big)^3 e^{-r/400} -0.2, 0\Bigg],
\end{equation}
where $\rho_{\mathrm{max}}$ is the maximum plasma density in the FM torus. When $\phi_{b}\ll\phi_{b,\text{crit}}$, we attain a standard and normal evolution (SANE) disk (\citealt{narayan_sane_2012, sadowski_sane_2013}). $A_{\phi}(r,\theta)$ for SANE disks in our simulations is given by,
\begin{equation}\label{eqn: sane_four_potential}
    A_{\phi} = \text{max}\Bigg[\frac{\rho}{\rho_{\text{max}}}-0.2, 0\Bigg].
\end{equation}
A poloidal slice of the initial conditions for one of our simulations (SANE $a_{*}=+0.94$) is shown in Figure \ref{fig:grid_init}.

\subsection{Grid geometry}\label{subsec:grid_geometry}

We integrate the governing equations (Equations \ref{eqn:ideal_mhd_equations_conservation_form}) in Kerr spacetime with a modified version of the Kerr-Schild (KS) coordinate system dubbed funky modified Kerr-Schild (FMKS) coordinate system $x^{\mu}=(x^0, x^1, x^2, x^3)$. The modification is two-fold: 
\begin{itemize}
    \item We use an exponential radial coordinate $x^{1}$ such that $r = \text{exp}(x^1)$, which increases the density of grid zones close to the event horizon.
    \item The polar coordinate is modified to increase the density of grid zones close to the midplane ($\theta=\pi/2$) and to widen the grid zones close to the poles  at small radii. The former is done to capture accretion disk physics with higher effective resolution while the latter increases the timestep in our simulations.
\end{itemize}
Figure \ref{fig:grid_init} plots the gridlines for one of the simulations considered in this work. For the exact coordinate transformations see Appendix F in \cite{Wong_2022_patoka} (also see section 4.2.1 in \citealt{prather_2022_thesis}).

\subsection{Failure modes}\label{subsec:failure_modes}

GRMHD codes are not robust in regions where the magnetic energy density exceeds the fluid energy density. Truncation errors in the evolution of the conserved quantities $\boldsymbol{U}$ can produce unphysical values of fluid quantities $\boldsymbol{P}$. In such scenarios GRMHD codes typically impose ``floors'' on $\rho$ and $u$ to ensure non-negative values, and limit the fluid Lorentz factor $\Gamma$ to avoid superluminal velocities. For the suite of simulations presented in this work, we apply the floors in the normal observer frame \citep{mckinney_general_2012} which in turn necessitates an additional $\boldsymbol{U}\rightarrow\boldsymbol{P}$ operation for the floored grid zones \footnote{We use the ``$1D_{W}$'' scheme \citep{noble_primitive_2006} as outlined in \cite{mignone_equation_2007}. This involves a 1D Newton-Raphson solve for the fluid primitives. The magnetic field primitives can be recovered analytically as they differ from the conserved variables by a multiplicative factor of $\sqrt{-g}$.}. In the event this inversion operation is unsuccessful, we average over neighboring zones that did manage to invert successfully (an operation hereafter referred to as a ``fixup''). The details of the application and values of floors and fixups in the library presented in this work are discussed in Appendix \ref{appendix:failure_modes}.

\subsection{Simulation suite}

The \v3 library consists of 10 GRMHD simulations run out to $30,000\hspace{0.1cm}t_{g}$ ($t_{g}\equiv GM/c^{3}$ is the light-crossing time) that span the $(\phi_b,a_{*})$ parameter space (hereafter, referred to as the \texttt{v3} library), where $a_{*}\equiv Jc/GM^2$ is the dimensionless black hole spin. Along the $\phi_b$ axis, we have the MAD and SANE accretion states and along the $a_{*}$ axis we consider 5 data points: $0, \pm1/2, \pm15/16$. To ensure consistency with other GRMHD simulations used in \citetalias{SgrAPaperV} we set $\hat{\gamma}=4/3$. We list the model parameters for the simulation library in Table \ref{table:grmhd_models}.

\setlength{\tabcolsep}{12pt}
\begin{deluxetable*}{ lrccclccc }
\tablecaption{GRMHD Simulation Suite} \label{table:grmhd_models}
\tablehead{
\colhead{Flux} & 
\colhead{$a_*$} &
\colhead{$r_{\text{in}}$ $(r_{g})$} &
\colhead{$r_{\text{max}}$ $(r_{g})$} &
\colhead{$r_{\text{out}}$ $(r_{g})$} &
\colhead{$\hat{\gamma}$} &
\colhead{Resolution} &
\colhead{Duration ($t_g$)}}
\startdata
MAD & $-0.94$ & $20$ & $41$ & $1000$ & 4/3 & 288x128x128 & 30,000  \\
MAD & $-0.5$ & $20$ & $41$ & $1000$ & 4/3 & 288x128x128 & 30,000  \\
MAD & $0$ & $20$ & $41$ & $1000$ & 4/3 & 288x128x128 & 30,000  \\
MAD & $0.5$ & $20$ & $41$ & $1000$ & 4/3 & 288x128x128 & 30,000  \\
MAD & $+0.94$ & $20$ & $41$ & $1000$ & 4/3 & 288x128x128 & 30,000   \\
\hline
SANE & $-0.94$ & $10$ & $20$ & $1000$ & 4/3 & 288x128x128 & 30,000  \\
SANE & $-0.5$ & $10$ & $20$ & $1000$ & 4/3 & 288x128x128 & 30,000  \\
SANE & $0$ & $10$ & $20$ & $1000$ & 4/3 & 288x128x128 & 30,000  \\
SANE & $0.5$ & $10$ & $20$ & $1000$ & 4/3 & 288x128x128 & 30,000  \\
SANE & $+0.94$ & $10$ & $20$ & $1000$ & 4/3 & 288x128x128 & 30,000  
\enddata
\tablecomments{Simulation parameters: Flux specifies the amount of magnetic flux threading the event horizon and can be MAD or SANE (see section \ref{subsec:initial_conditions}), $a_*$ is the dimensionless black hole spin, $\hat{\gamma}$ is the fluid adiabatic index, $r_{\text{in}}$ and $r_{\text{max}}$ are the inner and pressure maximum radii of the FM torus, and $r_{\text{out}}$ is the radial outer boundary of the simulation domain, all in units of $r_{g}\equiv~GM/c^2$. Resolution denotes the number of grid zones along each direction like $N_{r}\times N_{\theta}\times N_{\phi}$. Duration is the total duration of the simulation in units of $t_{g}$.}
\end{deluxetable*}

\subsection{Resolving the MRI}

The MRI facilitates the outward transport of angular momentum in a differentially rotating accretion disk with a weak magnetic field. The linear instability amplifies the seed magnetic field and causes a breakdown of the laminar flow into turbulence. The turbulent shear stress that arises acts as the primary channel for radial transport of angular momentum in the SANE simulations. 

We check if we are able to resolve the fastest growing mode of the MRI by computing ``MRI quality factors'' ($Q_{\text{MRI}}^{\theta}$ and $Q_{\text{MRI}}^{\phi}$; \citealt{sano_angular_2004,noble_dependence_2010,hawley_assessing_2011,hawley_testing_2013,narayan_sane_2012,porth_cc_2019}). These indicate the number of zones within a single wavelength of the fastest growing mode along the polar and azimuthal direction respectively. In the SANE simulations we find $Q_{\text{MRI}}^{\theta}\sim 5-10$ and $Q_{\text{MRI}}^{\phi}\sim 12-16$ \footnote{These are time- and azimuth-averaged values that were computed in the midplane of our simulation domain, from the inner radial boundary out to $r=50r_{g}$.}. While this is below the nominal values of $Q_{\text{MRI}}^{\theta,\phi}\sim 10,20$ suggested by \citet{hawley_assessing_2011,narayan_sane_2012}, we do achieve the minimum bound of $Q_{\text{MRI}}^{\theta}\geq 6$ prescribed by \citet{sano_angular_2004} in most of our simulations.
\section{Diagnostics and results}
\label{sec:discussions_results}

\subsection{Radial fluxes at the horizon}
\label{sec:fluxes}

In this subsection we study trends in the at-horizon radial fluxes in our simulation set.

\subsubsection{Time series}

\begin{figure*}
\centering
\includegraphics[,width=\linewidth]{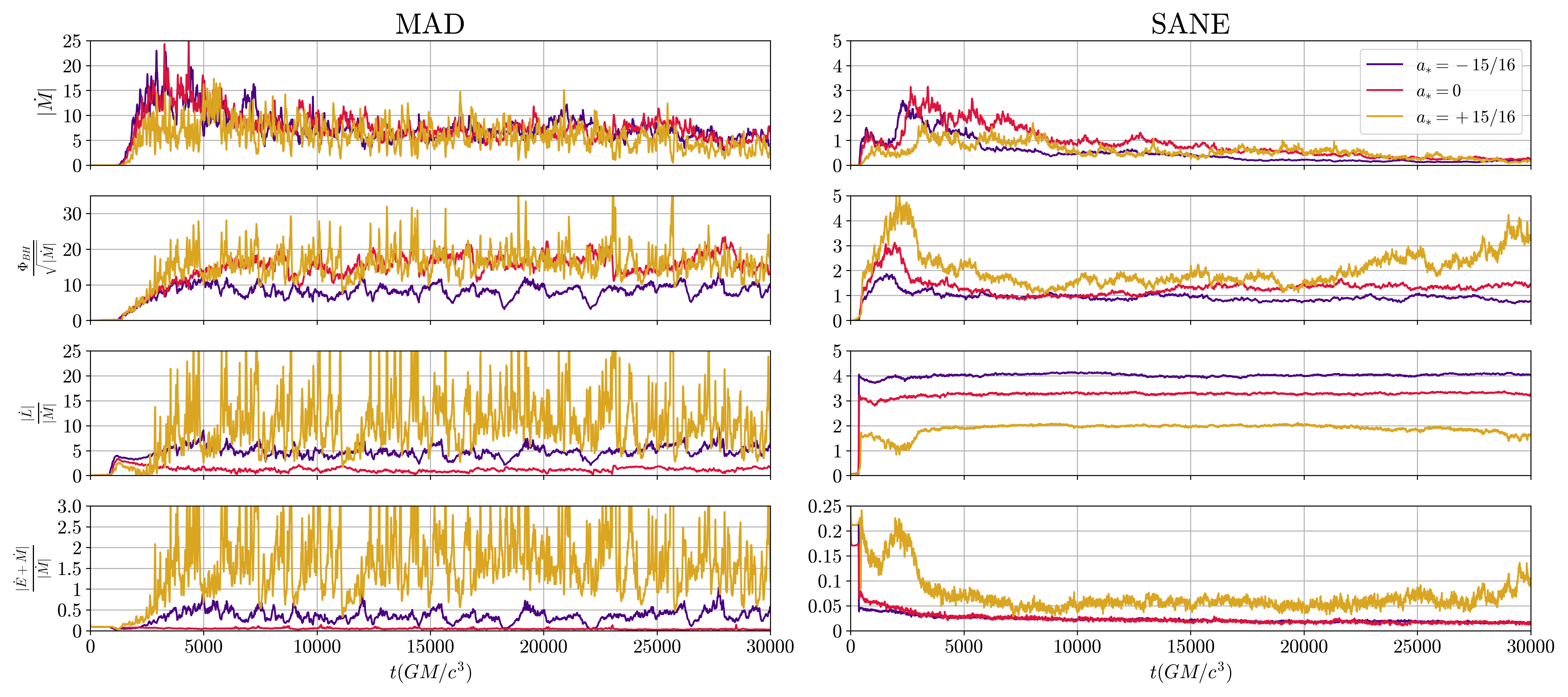}
\caption{Time series of the horizon-penetrating fluxes for six models: $a_{*}=\{-15/16,0,+15/16\}$, MAD and SANE. Left column: MAD simulations. Right column: SANE simulations. Top row: Rest-mass accretion rate. Second row: Dimensionless magnetic flux. Third row: Normalized, absolute angular momentum flux. Bottom row: Normalized absolute energy flux with the contribution from rest mass subtracted. $a_{*}=-15/16$ simulations are plotted in indigo, $a_{*}=0$ simulations in crimson, and $a_{*}=+15/16$ in goldenrod.}
\label{fig:fiducial_flux_time_series}
\end{figure*}

We evaluate radial fluxes of the rest-mass, magnetic field, total angular momentum and energy at the event horizon ($r_{\text{eh}}\equiv1+\sqrt{1-a_{*}^2}$) \footnote{In MAD models we measure $\dot{M}$, $\dot{L}$ and $\dot{E}$ at $r=5~r_g$. This is done to avoid the effect of density floors that tend to produce a noticeable difference in the MAD simulations (see relative difference in $\dot{M}$ in Table \ref{table:inflow_summary_statistics})},
\begin{align}
    \dot{M} &= \int_{\phi}\int_{\theta}(-\rho u^r)\sqrt{-g}\:d\theta\:d\phi, \label{eqn: accretion_rate_definition}\\
    \Phi_{\text{BH}} &= \frac{1}{2} \int_{\phi}\int_{\theta}\vert B^r\vert\sqrt{-g}\:d\theta\:d\phi, \label{eqn: magnetic_flux_definition}\\
    \dot{L} &= \int_{\phi}\int_{\theta}T^{r}_{\phi}\sqrt{-g}\:d\theta\:d\phi, \label{eqn: angular_momentum_flux_definition}\\
    \dot{E} &= \int_{\phi}\int_{\theta}(-T^{r}_{t})\sqrt{-g}\:d\theta\:d\phi, \label{eqn: energy_flux_definition}
\end{align}
where the integrals over ($\theta,\phi$) are over the entire shell.\footnote{The equations here are expressed in Kerr-Schild coordinates for clarity, but we perform the integration in FMKS coordinates.} $\dot{M}$ is the \textit{inward} rate of matter. Equation \ref{eqn: magnetic_flux_definition} counts the \textit{total} number of magnetic field lines threading the horizon, irrespective of the direction, and then halves to account for just one hemisphere. In Appendix \ref{appendix:phi_bh} we discuss the caveats associated this definition. The components of $T^{\mu}_{\nu}$ corresponding to angular momentum and energy are obtained by contracting it with the respective Killing vector fields, $\xi^{\mu}_{(t)}\equiv\partial_{t}$ and $\xi^{\mu}_{(\phi)}\equiv\partial_{\phi}$ (see Section \ref{sec:conserved_currents}). Note that $\dot{L}$ and $\dot{E}$ here describe the \textit{outflow} of angular momentum and energy.

Ideal GRMHD has no inherent length, time, or density scale and quantities can be scaled in accordance with the physical system being analyzed during post-processing. As a result we normalize the fluxes by appropriate powers of the accretion rate.

We plot the time series of these quantities for six of our models in Figure \ref{fig:fiducial_flux_time_series}. We see in the accretion rate plot (top panels) that following the initial rise in $\dot{M}$ as matter first crosses the event horizon, there is a decline, and ultimately $\dot{M}$ saturates to a quasi-steady value. The time at which this steady state is achieved varies from model-to-model, but by $15,000\hspace{0.1cm} t_{g}$ all models have attained this state. In the remainder of this paper we consider data in the range $t=[15, 30] \times 10^3 t_{g}$ when computing time-average of quantities. 

The SANE models tend to have a steady flow with smaller variations in all the radial fluxes. They are characterized by an almost uniform inflow of material across the horizon with the magnetic field playing a negligible role in the flow dynamics other than to transfer angular momentum outward. In MAD models the magnetic flux at the horizon saturates to its maximum value $\phi_{b,\text{crit}}$ soon after accretion commences. At this stage, the flow near the horizon is in a state of unstable equilibrium as the outward magnetic pressure counterpoises the inward fluid pressure. This is colloquially known as an ``arrested'' state. This is followed by a release of excess magnetic flux (dubbed as a ``flux eruption'' event; see \citealt{tchekhovskoy_efficient_2011,mckinney_general_2012,ripperda_flares_2022,davelaar_synchrotron_waves_boundary_2023,vos_flux_eruption_2024,zhang_mad_survey_2024} for a more detailed discussion on flux eruption events and its implications on flow dynamics and synthetic observables) and $\phi_{b}$ drops while $\dot{M}$ rises. These intermittent flux eruption events result in the larger fluctuations in the time series for the MAD models.

\subsubsection{Time-averaged}

\begin{figure*}
\centering
\includegraphics[,width=\linewidth]{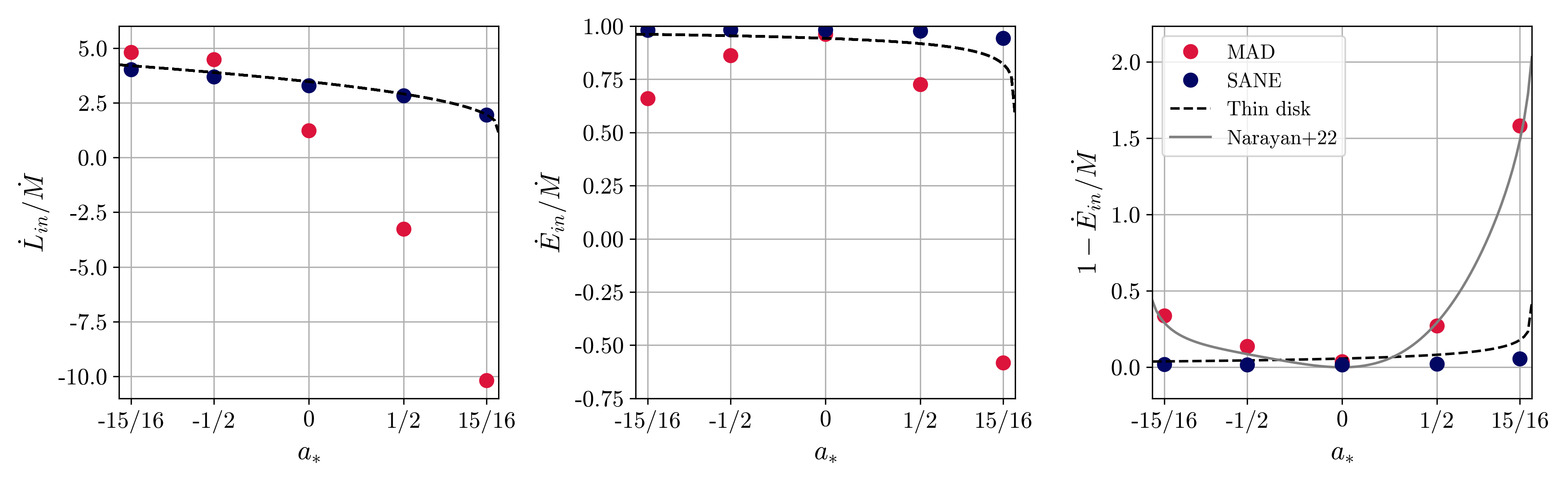}
\caption{Time-averaged radial fluxes as a function of black hole spin. The data points in red (blue) denote MAD (SANE) simulations. The dashed line plots the value expected for a thin disk at the ISCO. Left panel: Specific angular momentum flux. Middle panel: Specific energy flux. Right panel: Ratio of outflow power ($\dot{E} - \dot{M}$) to rest-mass accretion rate.}
\label{fig:flux_time_averaged}
\end{figure*}

\begin{figure*}
\centering
\includegraphics[,width=\linewidth]{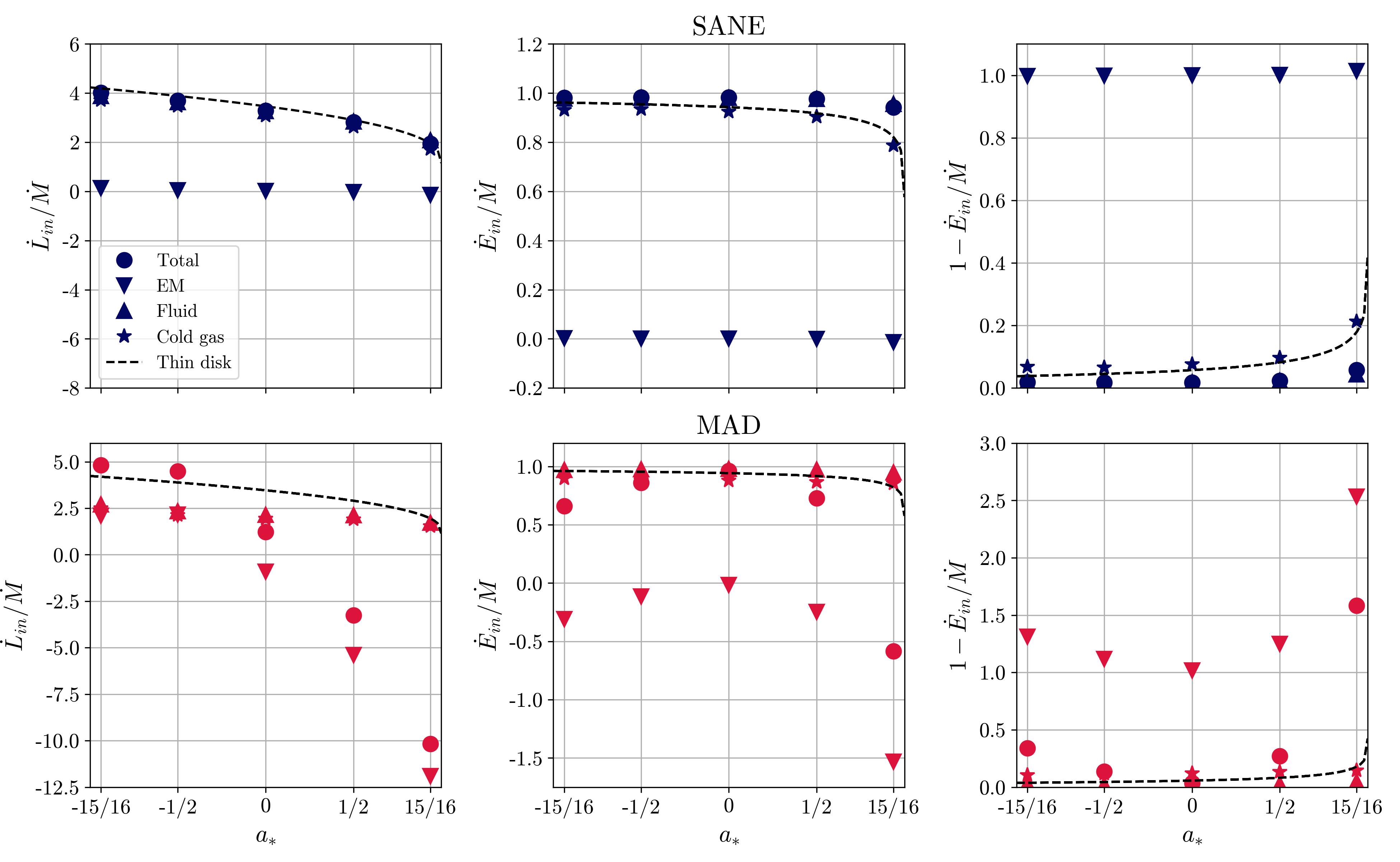}
\caption{Similar to Figure \ref{fig:flux_time_averaged} but now plotting the contribution of various components of the fluxes (See Equations \ref{eqn:ldot_em},\ref{eqn:edot_em}, \ref{eqn:ldot_fl}, \ref{eqn:edot_fl}, \ref{eqn:ldot_cg}, and \ref{eqn:edot_cg}). We see that magnetic fields have little-to-no contribution toward $\dot{L}$ and $\dot{E}$ in SANE models. In the MAD models the trend in total flux closely matches the trend observed in the electromagnetic components.}
\label{fig:flux_time_averaged_components}
\end{figure*}

We plot the time-averaged, normalized angular momentum ($l\equiv\dot{L}_{in}/\dot{M}$) and energy ($e\equiv\dot{E}_{in}/\dot{M}$) fluxes in the left and middle panels respectively in Figure \ref{fig:flux_time_averaged}. It is worth noting that (i) these are the \textit{inflowing} radial fluxes at the horizon, i.e., they differ by a negative sign from Equations \ref{eqn: angular_momentum_flux_definition}, \ref{eqn: energy_flux_definition}, and (ii) these are signed quantities unlike Figure \ref{fig:fiducial_flux_time_series} where we showed the absolute values of the fluxes.
The right panel in Figure \ref{fig:flux_time_averaged} plots the ``efficiency'' of the system. This is the ratio of mechanical power flowing out to the power flowing in due to fluid rest-mass accreting. Thin disk values at the innermost stable circular orbit $r_{\mathrm{isco}}$ \citep{shakura_sunyaev_1973, bardeen_rotating_bh_1972} are denoted by dashed lines. This corresponds to inflow of cold, unmagnetized gas in the equatorial plane.

The time-averaged specific angular momentum flux in SANE models closely follows the thin disk expectation. $l>0$ for all spins indicates a net inflow of angular momentum into the black hole. The $a_{*}\leq 0$ MAD models also exhibit a net transfer of angular momentum into the black hole from the surroundings. However, the prograde MAD models possess a powerful jet, with highly collimated magnetic field lines anchored at the horizon. These field lines remove angular momentum from the black hole and result in a net outflow (see Section \ref{sec:spinup}).

In the SANE simulations, most of the energy is advected into the black hole, and the deviation from the thin disk estimate grows as $a_{*}$ increases. We explore this in detail in the following paragraphs where we break down the contributions to these fluxes (Figure \ref{fig:flux_time_averaged_components}). MAD models generally exhibit substantially higher efficiencies with prograde models being more efficient than their retrograde counterparts. This aligns with results of \cite{tchekhovskoy_prograde_vs_retrograde_2012,tchekhovskoy_general_2012,narayan_jets_2022}.

Figure \ref{fig:flux_time_averaged_components} deconstructs the radial fluxes plotted in Figure \ref{fig:flux_time_averaged} into its electromagnetic,
\begin{align}
    l^{(\text{EM})} &= \frac{\langle -b^{2}u^{r}u_{\phi} + b^{r}b_{\phi} \rangle_{t}}{\langle-\rho u^{r}\rangle_{t}}, \label{eqn:ldot_em}\\
    e^{(\text{EM})} &= \frac{\langle b^{2}u^{r}u_{t} - b^{r}b_{t} \rangle_{t}}{\langle-\rho u^{r}\rangle_{t}}, \label{eqn:edot_em}
\end{align}
and fluid,
\begin{align}
    l^{(\text{fl})} &= \frac{\langle -(\rho + u + p_{g})u^{r}u_{\phi} \rangle_{t}}{\langle-\rho u^{r}\rangle_{t}}, \label{eqn:ldot_fl}\\
    e^{(\text{fl})} &= \frac{\langle (\rho + u + p_{g})u^{r}u_{t} \rangle_{t}}{\langle-\rho u^{r}\rangle_{t}}, \label{eqn:edot_fl}
\end{align}
components. $\langle\rangle_{t}$ denotes a time-averaging. Additionally, we plot the fraction of the fluid component corresponding to a cold gas,
\begin{align}
    l^{(\text{cg})} &= \frac{\langle -\rho u^{r}u_{\phi} \rangle_{t}}{\langle-\rho u^{r}\rangle_{t}}, \label{eqn:ldot_cg}\\
    e^{(\text{cg})} &= \frac{\langle \rho u^{r}u_{t} \rangle_{t}}{\langle-\rho u^{r}\rangle_{t}}. \label{eqn:edot_cg}
\end{align}

In SANE models, as one might expect, the electromagnetic components of the normalized fluxes are approximately zero due to the subdominant role of the magnetic fields. For the specific angular momentum flux, the cold gas component nearly accounts for the entire fluid contribution. In the case of specific energy flux, the cold gas component closely resembles that of a thin disk. Additionally, we observe that as spin increases, the thermodynamic contribution of the fluid grows, consistent with Figure \ref{fig:disk_avg_radial_profiles}, which shows an increase in fluid temperature with higher $a_{*}$. 

In MAD simulations, the electromagnetic component dominates the diagnostic trends, except in the non-spinning case. The specific energy flux in the fluid sector is approximately evenly distributed between rest-mass energy and internal energy.

\subsubsection{Time variability of $\dot{M}$}

\begin{figure}
\centering
\includegraphics[,width=\linewidth]{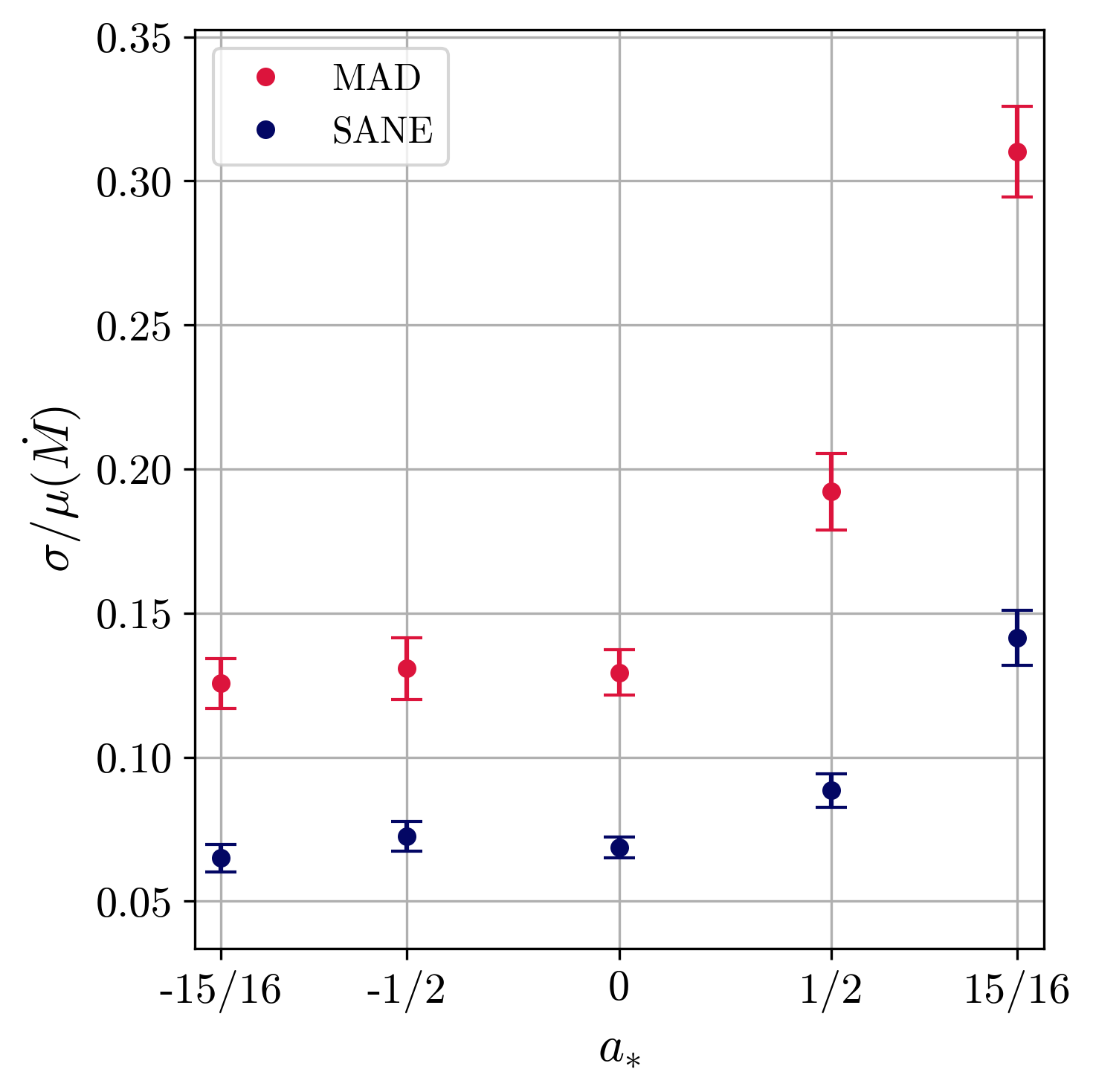}
\caption{3hr modulation index $M_{3}$ for the accretion rate $\dot{M}$ (defined as $M_{\Delta t}\equiv\sigma_{\Delta t}/\mu_{\Delta t}$; here $\Delta t=3$hr). To estimate the average $M_{3}$ (indicated by a marker for each simulation in this plot) for SgrA* (which has a characteristic timescale $t_{g}\sim20\hspace{0.1cm}$s), we extract as many independent 3 hour time segments from the $\dot{M}$ time series from t=15,000$t_{g}$ to t=30,000$t_{g}$ ($\sim27$) and compute the average sample mean. The errorbars plot the standard error of the sample means ($\sim\sigma_{\mathrm{M_{3}}}/\sqrt{27}$) and quantify the uncertainty in our measurement of $M_{3}$.}
\label{fig:mdot_variability}
\end{figure}

The primary motivation for this library was to provide a physical interpretation of the 2017 EHT observation of the galactic center \citepalias{SgrAPaperV} in conjunction with historical multifrequency observations of \sgra. In total, 11 heterogeneous constraints were applied to infer properties of the source. To minimize the possibility of wrongly rejecting a model, the GRMHD simulations were run out for a longer duration compared to previous work \citep{Wong_2022_patoka} to generate a larger set of independent samples of synthetic observations. This is particularly important when assessing model feasibility based on the 230GHz light curve variability (see the discussion in \citetalias{SgrAPaperV} and \citealt{wielgus_mm_lightcurves_2022}).

In this subsection we study the time variability in $\dot{M}$. While this does not directly translate to variability in the image-integrated flux density \footnote{The synchrotron emissivity for a population of relativistic thermal electrons is a function of the electron number density, magnetic field strength, electron temperature, and frequency \citep{leung_mbs_emission_absorption_2011}. Additionally, geometric effects such as gravitational lensing and Doppler beaming can alter the horizon-scale image and, in-turn, change the observed flux density.}, it can be a reasonable proxy given that most of the 230GHz compact flux observed by the EHT arises close to the event horizon (see, e.g., \citealt{porth_cc_2019} or Figure 5 in \citealt{Wong_2022_patoka}).

In Figure  \ref{fig:mdot_variability} we present the time variability in $\dot{M}$ for \sgra. The average and standard error of the sample means of the modulation index (defined as $M_{\Delta t}\equiv\sigma_{\Delta t}/\mu_{\Delta t}$ for a time duration $\Delta t$) are computed in a manner similar to the variability analysis in \citetalias{SgrAPaperV}. Independent 3 hour segments, corresponding to the typical decorrelation timescale in our GRMHD simulations (see \citealt{wielgus_mm_lightcurves_2022}), are extracted from the $\dot{M}$ time series starting at t=15,000$t_{g}$, and we calculate the sample mean $M_{3}$ ($\Delta t=3$ hours, $\sim 530~t_g$ for \sgra). The errorbars reflect our uncertainty in measuring the average $M_{3}$, which is directly proportional to the standard deviation of each sample, and inversely proportional to the number of independent samples. Two clear trends emerge: (i) MAD models exhibit significantly more variability than their SANE counterparts with the same black hole spin, a finding also seen in
studies of light curve variability (see the discussion on variability in \citetalias{SgrAPaperV}), and (ii) $M_{3}$ is higher for models with $a_{*}>0$, while retrograde models with the same magnetization state (MAD or SANE) show comparable $M_{3}$ values.

\subsection{Inflow equilibrium}
\label{sec:inflow_equilibrium}

We initialize our simulations with a Fishbone-Moncrief torus, a solution to the relativistic Euler equations in a stationary spacetime, and perturb the internal energy to seed instabilities, such as the MRI, which drive accretion (see Section \ref{subsec:initial_conditions}). The evolution of the accretion disk can be sensitive to the initial conditions. To reduce their influence on the GRMHD data products, simulations must be run out for longer periods, allowing the system to reach a steady state ``inflow equilibrium'' at larger radii. In this section we assess the radial extent over which our simulations achieve inflow equilibrium using two diagnostics.

First, we examine the time-averaged accretion rate as a function of radius $\dot{M}(r)$. As the fluid orbits the black hole, the velocity gradient induces a shear force between neighboring fluid annuli, generating viscous torques that transfer angular momentum outward and drive inflow (see, e.g., \citealt{lynden-bell_agn_1969, shakura_sunyaev_1973, novikov_thorne_1973, page_thorne_1974, lynden_bell_pringle_viscous_discs_1974, pringle_accretion_review_1981, frank_king_raine_accretion_book_1985, papaloizou_lin_review_1995}). In steady-state accretion disk theory, $\dot{M}$ is constant and independent of radius. In our simulations, however, we do not expect $\dot{M}$ to remain constant over time due to the finite reservoir of material, which depletes through both accretion into the black hole and outflows \footnote{There are also small-scale turbulent fluctuations, but time-averaging is expected to smooth these out.}. Nevertheless, as the simulation progresses, we expect the $\dot{M}$ profile to attain a constant value. To estimate an inflow equilibrium radius $r_{\mathrm{eq}}$ from $\dot{M}(r)$ profiles, we adopt the criterion from \citet{white_long_duration_2020}, where the inflow equilibrium radius is defined as the radius at which $\dot{M}$ drops to $e^{-1/2}$ of its value at the horizon ($5\hspace{0.1cm}r_g$ in the case of MAD simulations). The $r_{\mathrm{eq}}$ values for our simulations are listed in Table \ref{table:inflow_summary_statistics}.



\setlength{\tabcolsep}{5pt}
\begin{deluxetable}{ lrcccc }
\tablecaption{Inflow equilibrium and accretion rate statistics} \label{table:inflow_summary_statistics}
\tablehead{
\colhead{Flux} & 
\colhead{$a_*$} &
\colhead{$\dot{M}$ difference (\%)} &
\colhead{$r_{\text{eq}}(\dot{M})$} &
\colhead{$r_{\text{eq}}(t_{\mathrm{in}})$}
}
\startdata
MAD & -15/16 & 6.09 & 64.00 & 50.65 \\ 
MAD & -1/2 & 7.35 & 65.55 & 51.32 \\ 
MAD & 0 & 8.56 & 70.59 & 49.65 \\ 
MAD & +1/2 & 12.15 & 59.97 & 46.96 \\ 
MAD & +15/16 & 20.40 & 45.06 & 45.06 \\ 
SANE & -15/16 & 0.13 & 30.99 & 26.31 \\ 
SANE & -1/2 & 0.17 & 30.09 & 26.33 \\ 
SANE & 0 & 0.39 & 31.98 & 27.41 \\ 
SANE & +1/2 & 0.89 & 28.15 & 22.54 \\ 
SANE & +15/16 & 1.77 & 18.52 & 17.67 \\ 
\enddata
\tablecomments{The first two columns are identical to Table \ref{table:grmhd_models}. The third column mentions the relative difference in $\dot{M}$ when measured at $r_{\text{eh}}$ vs 5$r_g$, $r_{\text{eq}}(\dot{M})$ and $r_{\text{eq}}(t_{\mathrm{in}})$ are the radial extent to which the flow has achieved steady state as  measured from $\dot{M}(r)$ and $t_{\mathrm{in}}(r)$ respectively.}
\end{deluxetable}

\begin{figure}
\centering
\includegraphics[,width=\linewidth]{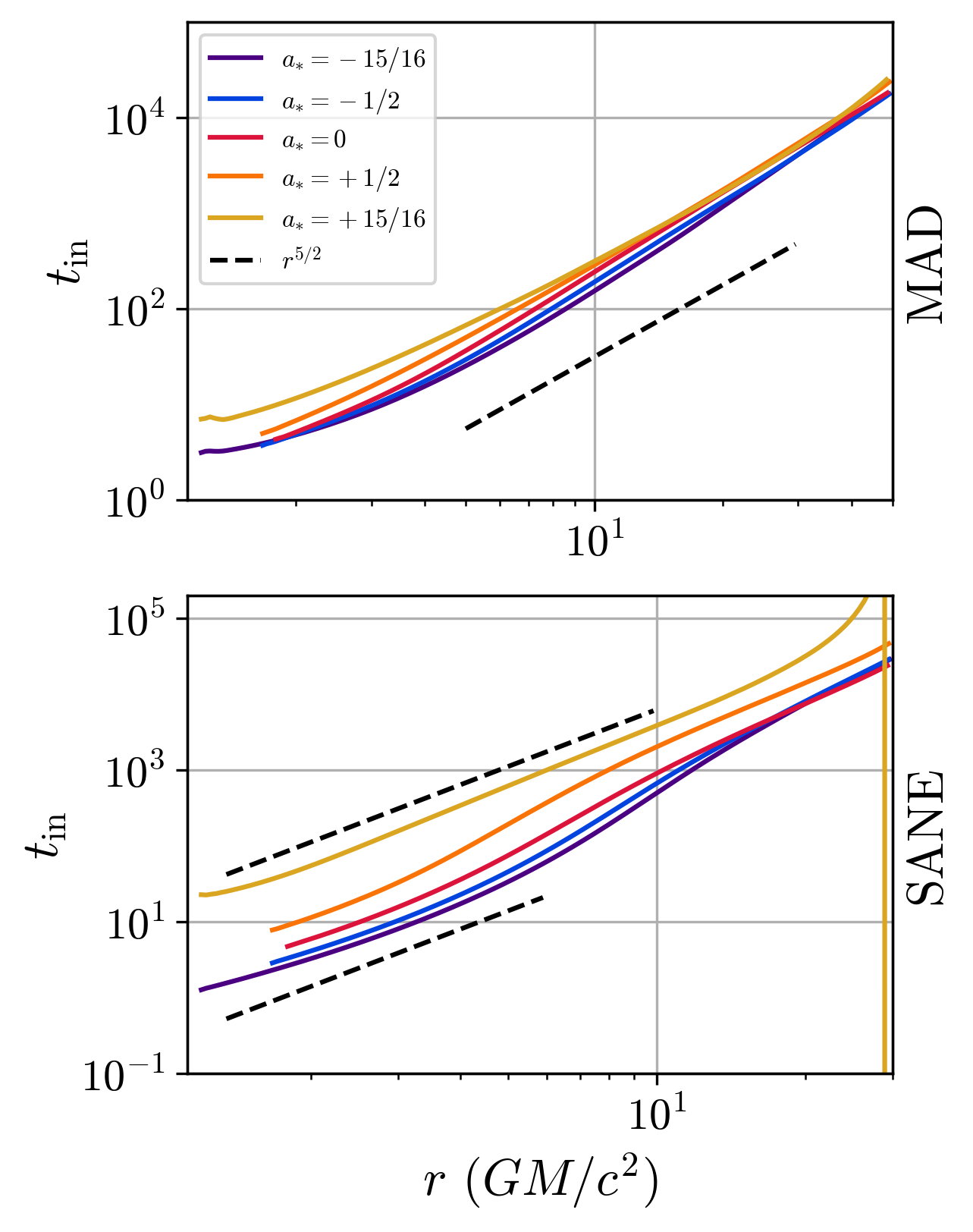}
\caption{Time-averaged, density-weighted radial profiles of $t_{\mathrm{in}}$ for the \v3 library. The dashed lines are power laws that approximately trace $t_{\mathrm{in}}$.}
\label{fig:tvisc_spins}
\end{figure}

Second, we evaluate the inflow timescale,
\begin{equation}
\label{eqn:viscous_timescale}
    t_{\mathrm{in}}(r) \equiv \frac{r}{\langle-v^{r}\rangle},
\end{equation}
where $v^{r}$ is the radial component of the spatial fluid velocity measured by the normal observer, $v^{i} = u^{i} / u^{t}$. This is often referred to as the viscous timescale \citep{penna_thin_disc_2010, narayan_sane_2012} as it represents the time required for viscous diffusion at a given radius $r$. Figure \ref{fig:tvisc_spins} shows the time-averaged $t_{\mathrm{in}}$ as a function of radius. As expected, the inflow time increases with radius. The MAD simulations exhibit lower $t_{\mathrm{in}}$ at the same radii compared to SANE simulations, indicating a higher inward radial velocity, consistent with findings of \cite{narayan_sane_2012}. Notably, in SANE simulations, $t_{\mathrm{in}}$ increases with increasing $a_{*}$. The dashed lines represent powerlaws $\sim r^{5/2}$ that approximate $t_{\mathrm{in}}(r)$, which is steeper than the thin-disk expectation $t_{\mathrm{in}}\sim r/v^{r}\sim r^{2}/\nu\sim r/(\alpha c_{s}(H/r))\sim r^{3/2}/(\alpha(H/r)^2)$, assuming a constant $\alpha$. $c_s$ is the local sound speed, and $\nu$ is the coefficient of kinematic viscosity. We define $r_{\mathrm{eq}}$ as the radius where the inflow timescale matches the simulation run time \footnote{We consider the end of the simulation when evaluating the simulation run time.}. The final column in Table \ref{table:inflow_summary_statistics} lists $r_{\mathrm{eq}}$ computed using the inflow timescale.

\subsection{Time-averaged disk structure}
\label{sec:disk_structure_avg}

\begin{figure*}
\centering
\includegraphics[,width=\linewidth]{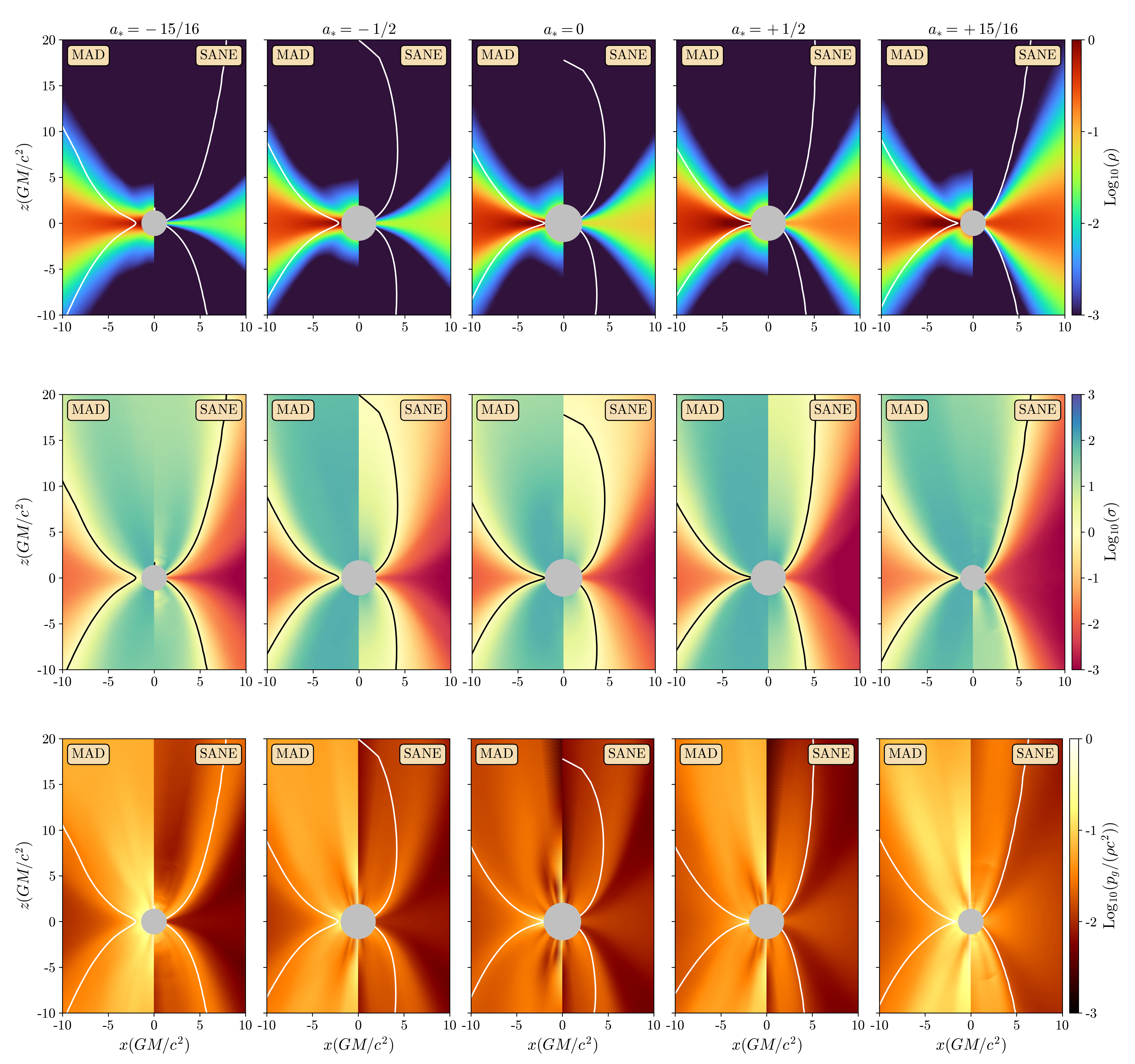}
\caption{Top to bottom: time- and azimuth-averaged poloidal plots of $\rho$, $\sigma\equiv b^{2}/\rho$, and $\Theta\equiv p_{g}/(\rho c^{2})$ respectively. Each column represents a different black hole spin. In each subplot, the left panel shows the MAD simulation, while the right panel shows the corresponding SANE simulation. Contours in each plot indicate $\sigma=1$. Note that for quantities defined as ratios, such as magnetization $\sigma$ and fluid temperature $\Theta$, the averages are taken separately, e.g., $\langle\sigma\rangle=\langle b^{2}\rangle/\langle\rho\rangle$.}
\label{fig:axisym}
\end{figure*}

It is instructive to look at the time- and azimuth-averaged quantities to infer broad characteristics of the flow. In Figure \ref{fig:axisym}, we present poloidal slices of the rest-mass density $\rho$, plasma magnetization $\sigma$, and fluid temperature $p_{g}/(\rho c^2)$, all averaged in time and azimuth. The contours indicate the $\sigma=1$ surface, which we use to delineate the magnetically dominated jet from the disk. MAD accretion disks are hotter than their SANE counterparts, a trend more clearly shown in Figure \ref{fig:disk_avg_radial_profiles} where we plot disk-averaged radial profiles. The halo of high density close to the black hole in the MAD simulations is due to numerical floors. We discuss this in greater detail in Appendix \ref{appendix:failure_modes}.

\begin{figure}
\centering
\includegraphics[,width=\linewidth]{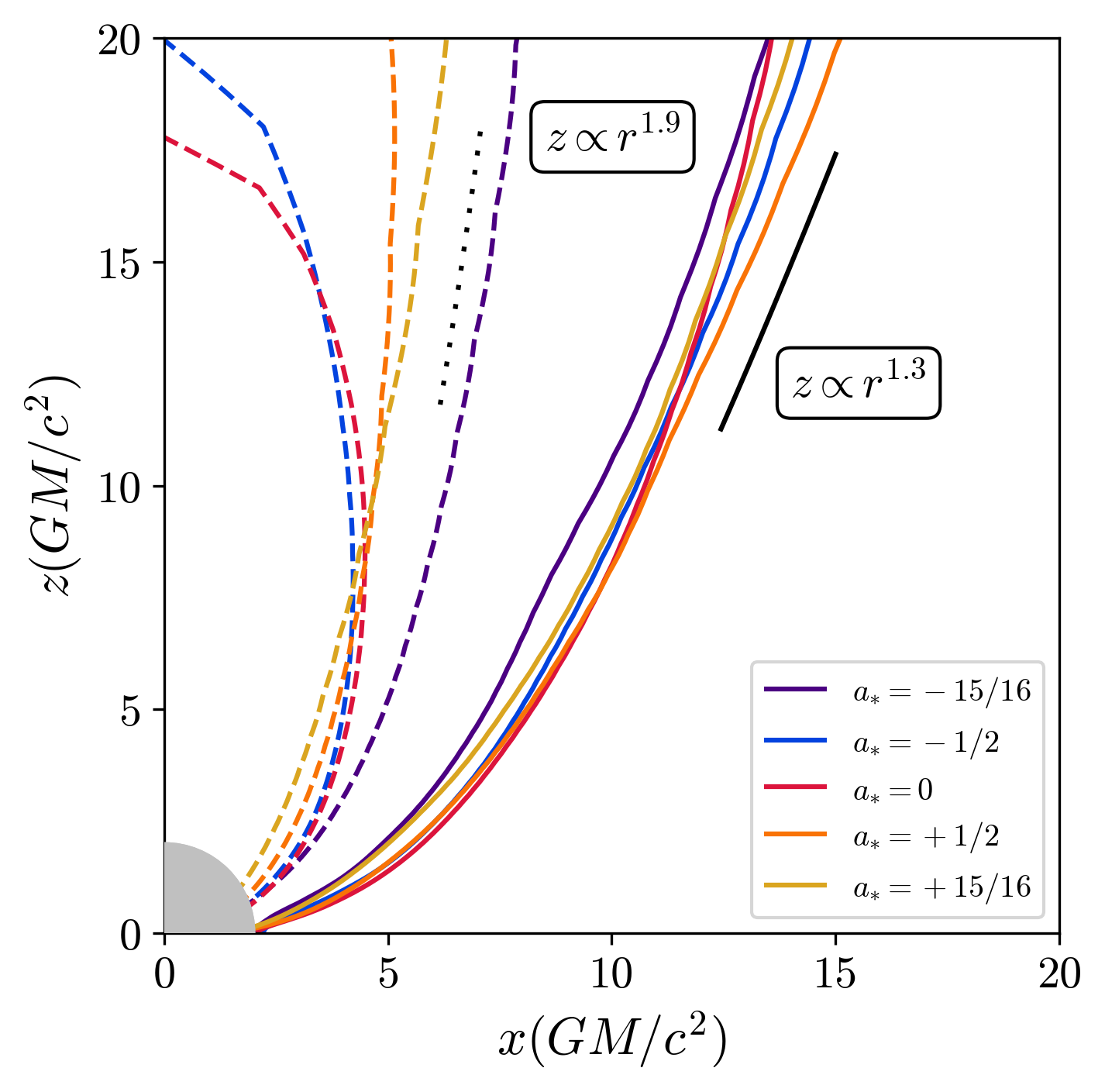}
\caption{Time- and azimuth-averaged ``jet profiles'' ($\sigma=1$ contours). The solid (dashed) lines represent MAD (SANE) models. The filled grey circle at the lower left corner denotes a zero spin black hole ($r_{\mathrm{eh}}=2~r_g$).}
\label{fig:jet_profiles}
\end{figure}

\begin{figure*}
\centering
\includegraphics[,width=\linewidth]{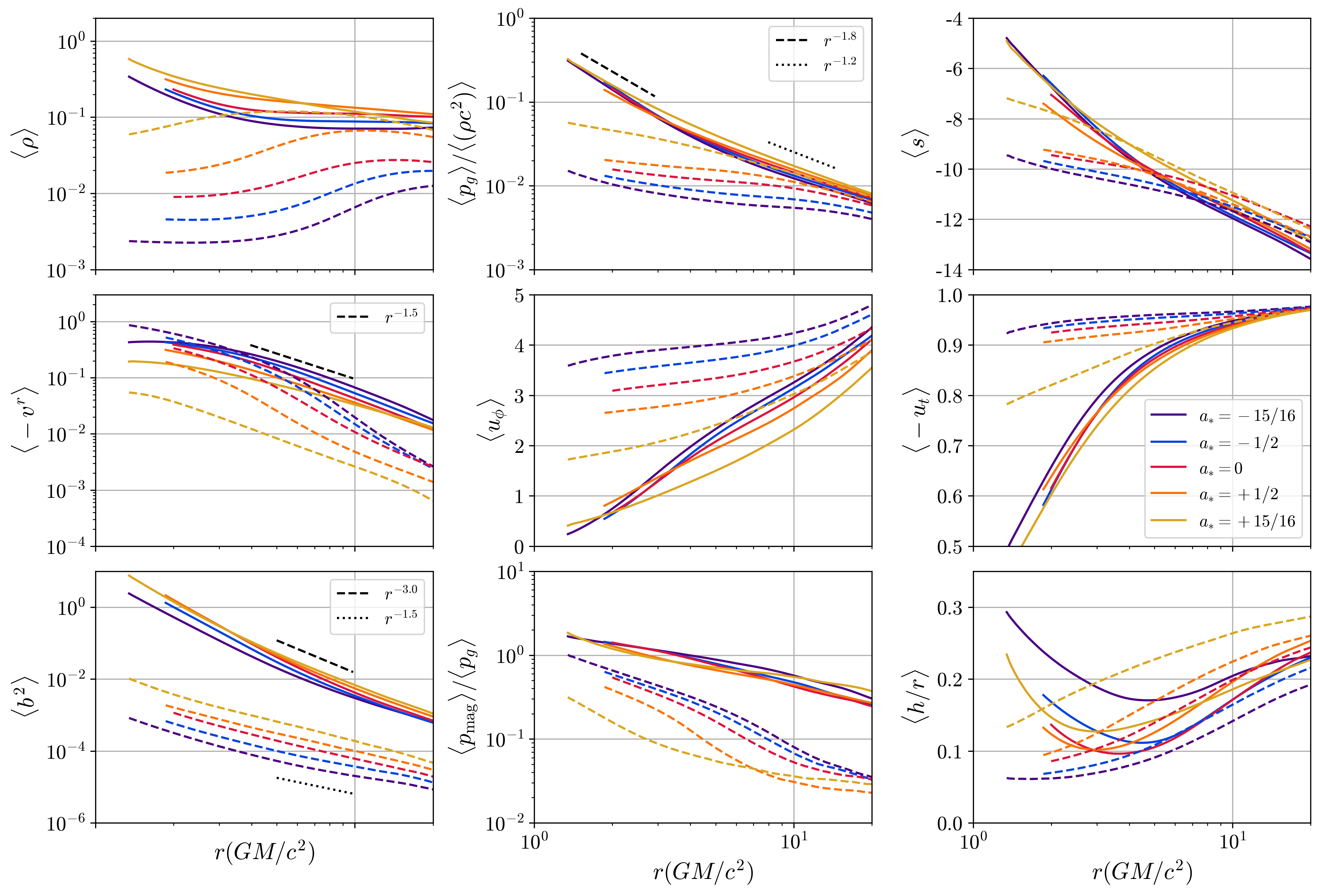}
\caption{Radial profiles of rest-mass density $\rho$ (in arbitrary code units), dimensionless fluid temperature $\Theta\equiv p_{g}/(\rho c^{2})$, fluid entropy $s$, radial velocity $v^r$, specific angular momentum $u_{\phi}$, specific energy $u_t$, magnetic field strength squared $b^2$, inverse plasma-beta $\beta_{p}^{-1}\equiv(b^{2}/2)/p_{g}$, and the density scale height $h/r$. The solid (dashed) lines represent MAD (SANE) simulations. Different colors correspond to different spins. Note that the units of rest-mass density in our simulations are arbitrary---a consequence of the scale-free nature of ideal GRMHD simulations---and therefore the absolute values of $\langle\rho\rangle$ have limited physical significance. One should instead focus on the structure of the radial profile.}
\label{fig:disk_avg_radial_profiles}
\end{figure*}

In Figure \ref{fig:jet_profiles} we present the jet profiles ($\sigma$ = 1 contours). The MAD models exhibit a wider jet compared to the corresponding SANE models. In the SANE $a_{*}=-0.5,0,+0.5$ simulations, the outflow is too weak to sustain a persistent jet, causing the $\sigma=1$ contour to loop back inward. Unlike \cite{narayan_jets_2022}, we do not observe a strong dependence of the jet profile on spin in the MAD models, possibly due to the shorter duration of our simulations. We also show powerlaws approximating the jet profiles near the black hole: $z\propto r^{1.3}$ for the MAD simulations and $z\propto r^{1.9}$ for the SANE simulations. These results are consistent with previous numerical studies \citep{mckinney_general_2006, narayan_jets_2022} and observational work on AGNs \citep{asada_m87_jet_profile_observations_2012}, which predict a parabolic jet profile within the inner $\sim10^{4}\hspace{0.1cm}r_{g}$.

Figure \ref{fig:disk_avg_radial_profiles} shows radial profiles of the rest-mass density $\rho$, dimensionless fluid temperature $p_{g}/(\rho c^2)$, fluid entropy $s\equiv(\hat{\gamma}-1)~\mathrm{log}(p_g/\rho^{\hat{\gamma}})$, the radial component of the fluid velocity $v^{r}$, the specific angular momentum $u_{\phi}$, specific energy $u_{t}$, the magnetic field strength squared $b^2$, the ratio of magnetic to gas pressure $b^2/(2p_{\mathrm{g}})$, and the density scale height $h/r$. We compute a density-weighted average,
\begin{equation}
\label{eqn:density_weighted_radial_profiles}
    \langle q(r)\rangle_{t,\phi}\equiv\frac{\int\int\int q(t,r,\theta,
    \phi)\cdot\rho(t,r,\theta,
    \phi)\cdot\sqrt{-g}\mathrm{d}t\mathrm{d}\theta\mathrm{d}\phi}{\int\int\int\rho(t,r,\theta,
    \phi)\cdot\sqrt{-g}\mathrm{d}t\mathrm{d}\theta\mathrm{d}\phi},
\end{equation}
for $p_g/\rho, s, v^r, u_{\phi}, u_{t}, b^2$ and $b^2/(2p_{\mathrm{g}})$ to highlight the behavior of these quantities in the accretion disc. The density profile is volume-averaged. For quantities that are ratios of independent variables, e.g., $b^2/(2p_{g})$, we compute the density-weighted average of each variable separately before calculating the ratio. The density scale height is computed as,
\begin{equation}
\label{eqn:density_scale_height}
    \frac{h}{r}(r)\equiv\frac{\int\int\int \rho(t,r,\theta,
    \phi)\vert\pi/2-\theta\vert\cdot\sqrt{-g}\mathrm{d}t\mathrm{d}\theta\mathrm{d}\phi}{\int\int\int\rho(t,r,\theta,
    \phi)\cdot\sqrt{-g}\mathrm{d}t\mathrm{d}\theta\mathrm{d}\phi}.
\end{equation}
We find the MAD models are approximately an order of magnitude hotter than the SANE simulations within the inner $\sim 10~r_g$, likely due to compressional heating. In MAD simulations, the magnetically dominated, wider jets constrain accretion to narrow channels. In the case of SANE simulations the temperature increases with increasing $a_{*}$.

The ratio of magnetic pressure to gas pressure is higher in MAD simulations. For MAD simulations, the radial fluid velocity seems to follow a power law $\langle v^r\rangle\propto r^{3/2}$, independent of spin. In contrast, $\langle v^r\rangle$ decreases with increasing $a_{*}$ in the inner $\sim10\hspace{0.1cm}r_{g}$ for SANE simulations. This is expected if the fluid begins to plunge within the ISCO due to insufficient centrifugal support. The average specific angular momentum $\langle u_{\phi}\rangle$ is a strong function of $a_{*}$ in SANE flows, with retrograde models transferring angular momentum to the black hole more efficiently, reaffirming the findings in Section \ref{sec:fluxes} and specifically Figure \ref{fig:flux_time_averaged_components}. However, in MAD models, the spin dependence is less pronounced. The density scale height $h/r$ is directly proportional to $a_{*}$ for SANE simulations. Notably, for the SANE models (i) the radial structure and (ii) the spin dependence of the average temperature, specific angular momentum, and scale height agree remarkably well with the semi-analytical model of \cite{gammie_popham_adaf_1998, popham_gammie_adaf_1998} which solves the height-integrated, axisymmetric equations of relativistic hydrodynamics with a causal prescription for the viscous stress-energy tensor.

\begin{figure}
\centering
\includegraphics[,width=\linewidth]{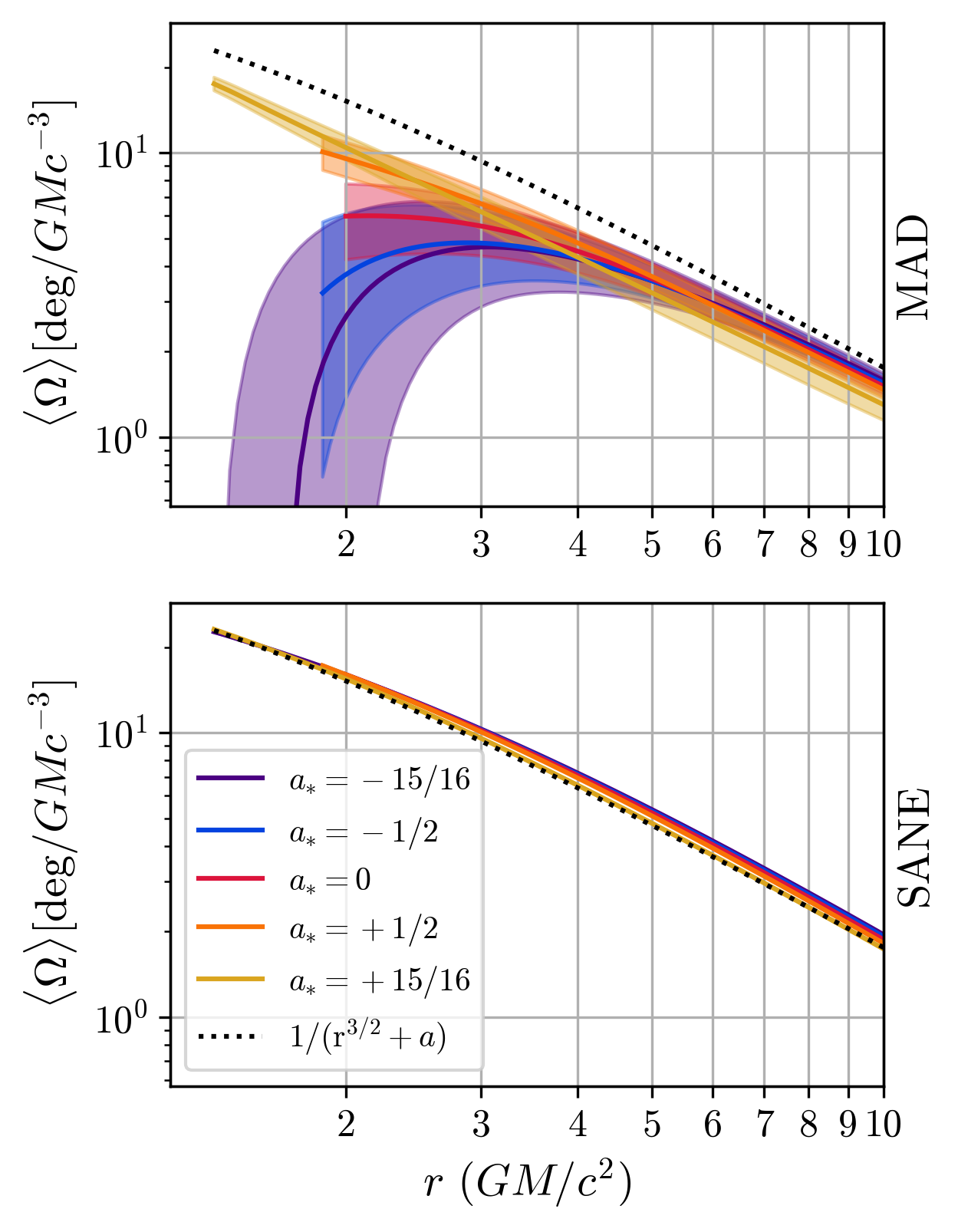}
\caption{Angular velocity  profiles $\langle\Omega\rangle\equiv\langle u^{\phi}\rangle/\langle u^{t}\rangle$ in Kerr-Schild coordinates in units of deg/GM/c$^3$. The solid line plots the time-average value and the shaded region plots one standard deviation. The dotted black line is the Keplerian fit $\Omega_{K}=(r^{3/2} + a)^{-1}$ for $a_{*}=+0.94$. The MAD rotational profiles are sub-Keplerian and have a significantly greater spread as compared to the SANE models. The SANE angular velocity agrees with the Keplerian expectation.}
\label{fig:disk_avg_rotational_profile}
\end{figure}

In Figure \ref{fig:disk_avg_rotational_profile} we plot the density-weighted average angular velocity $\langle\Omega\rangle\equiv\langle u^{\phi}\rangle/\langle u^{t}\rangle$. The MAD simulations exhibit sub-Keplerian behavior with significant variance in azimuthal fluid velocity. SANE flows closely follow the Keplerian profile and display a more organized angular motion. 

\subsection{Conserved currents}
\label{sec:conserved_currents}

\begin{figure*}
\centering
\includegraphics[,width=0.7\linewidth]{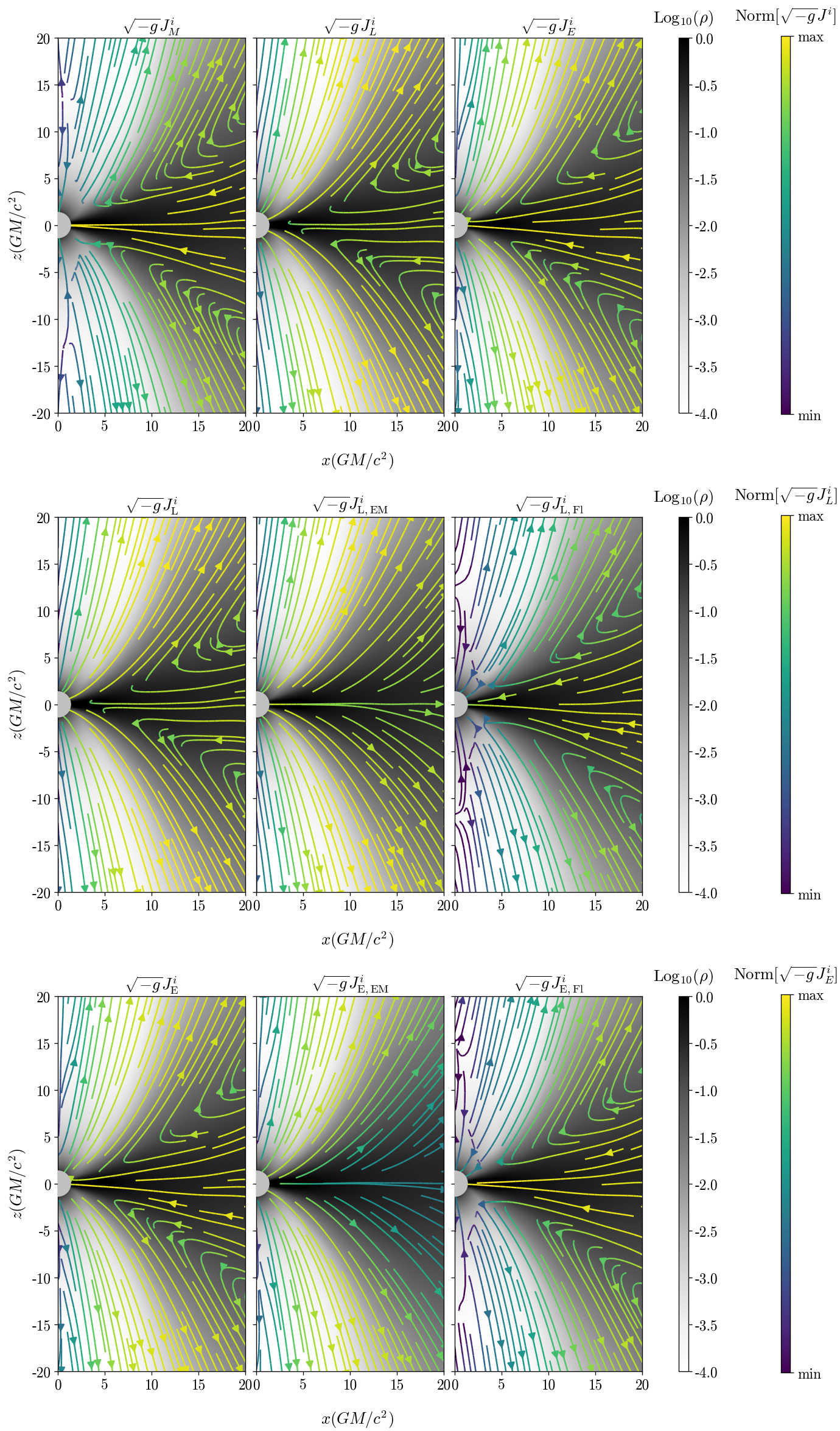}
\caption{Time- and azimuth-averaged poloidal slices of rest-mass density with averaged conserved currents overlaid for the MAD $a_{*}=+0.94$ model. In the top row we plot (left to right) the mass $\boldsymbol{\mathfrak{J}}_{M}\equiv\sqrt{-g}\rho u^{\mu}$, angular momentum $\boldsymbol{\mathfrak{J}}_{L}\equiv \sqrt{-g}T^{\mu}_{\phi}$, and energy $\boldsymbol{\mathfrak{J}}_{E}\equiv-\sqrt{-g}T^{\mu}_{t}$ currents. The middle row shows from left to right the total angular momentum current, the electromagnetic contribution, and the fluid contribution respectively. The bottom row is similar to the middle row except we plot the energy current. The arrows in the flow lines indicate the direction of the conserved currents and the colorscale indicates the magnitude plotted on a logscale since it can vary by several decades on a poloidal slice. The general morphology of the currents is similar, namely, strong inflow in the disk that transitions to an outflow at higher latitudes.}
\label{fig:conserved_currents_mad+0.94}
\end{figure*}

\begin{figure*}
\centering
\includegraphics[,width=0.7\linewidth]{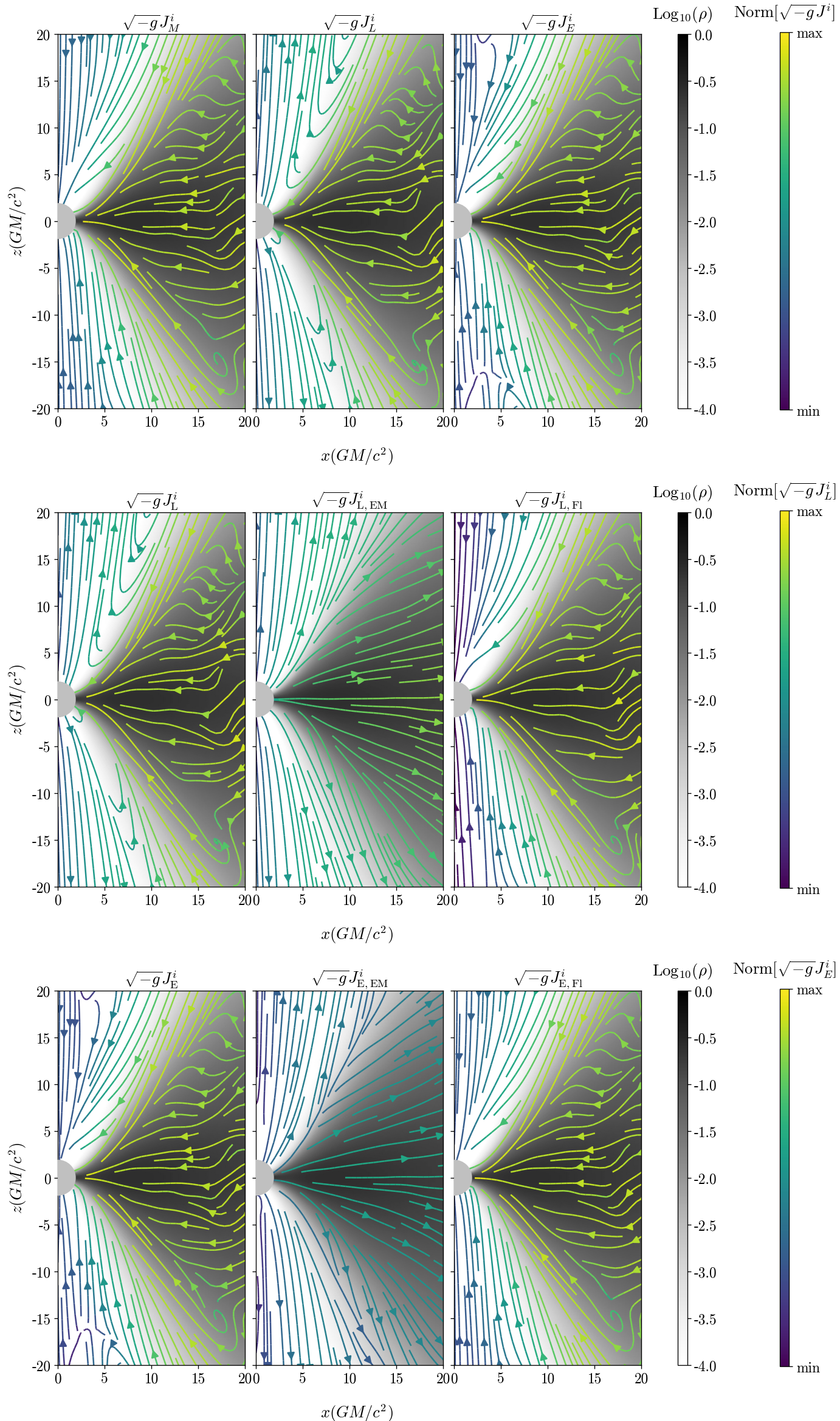}
\caption{Same as Figure \ref{fig:conserved_currents_mad+0.94} but for SANE $a_{*}=+0.5$}
\label{fig:conserved_currents_sane+0.5}
\end{figure*}

\begin{figure}
\centering
\includegraphics[,width=\linewidth]{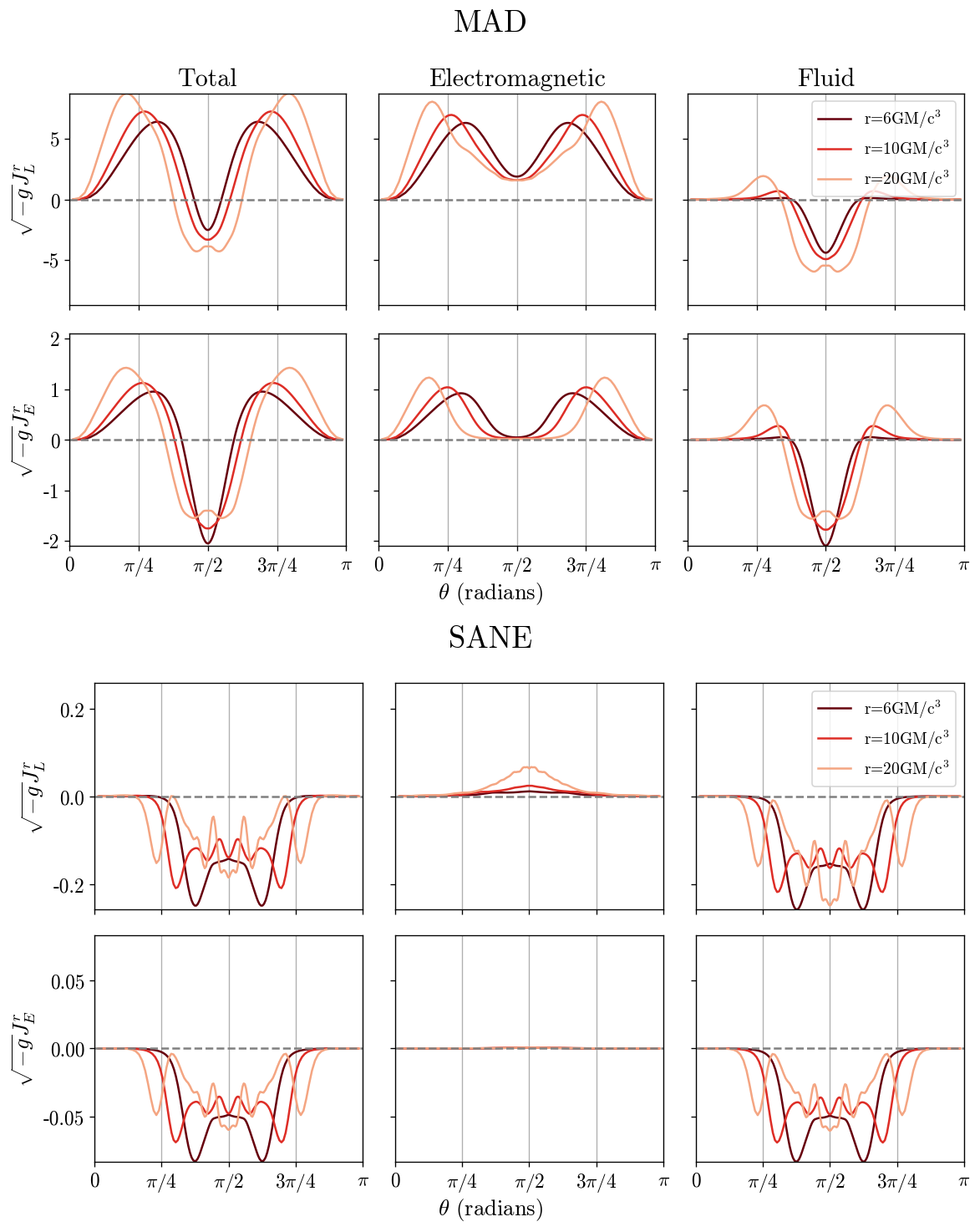}
\caption{$\boldsymbol{\mathfrak{J}}_{L}$ and $\boldsymbol{\mathfrak{J}}_{E}$ colatitude dependence at three radii $r=6,10,20~r_{g}$. The top (bottom) two rows corresponds to the simulation shown in Figure \ref{fig:conserved_currents_mad+0.94} (Figure \ref{fig:conserved_currents_sane+0.5}). For each simulation we show in the top (bottom) row the angular momentum (energy) current. The left column plots the \textit{total} current, while the middle and right columns plot the electromagnetic and fluid contribution respectively. In the MAD simulation we see outward transport of energy and angular momentum flux in the jet carried by the collimated magnetic fields; while the fluid stresses are inward and peak in the disk. The total current closely resembles the fluid sector profile for the SANE simulation with some outward electromagnetic angular momentum flux in the disk.}
\label{fig:conserved_currents_colatitude_radial_dependence}
\end{figure}

\begin{figure}
\centering
\includegraphics[,width=\linewidth]{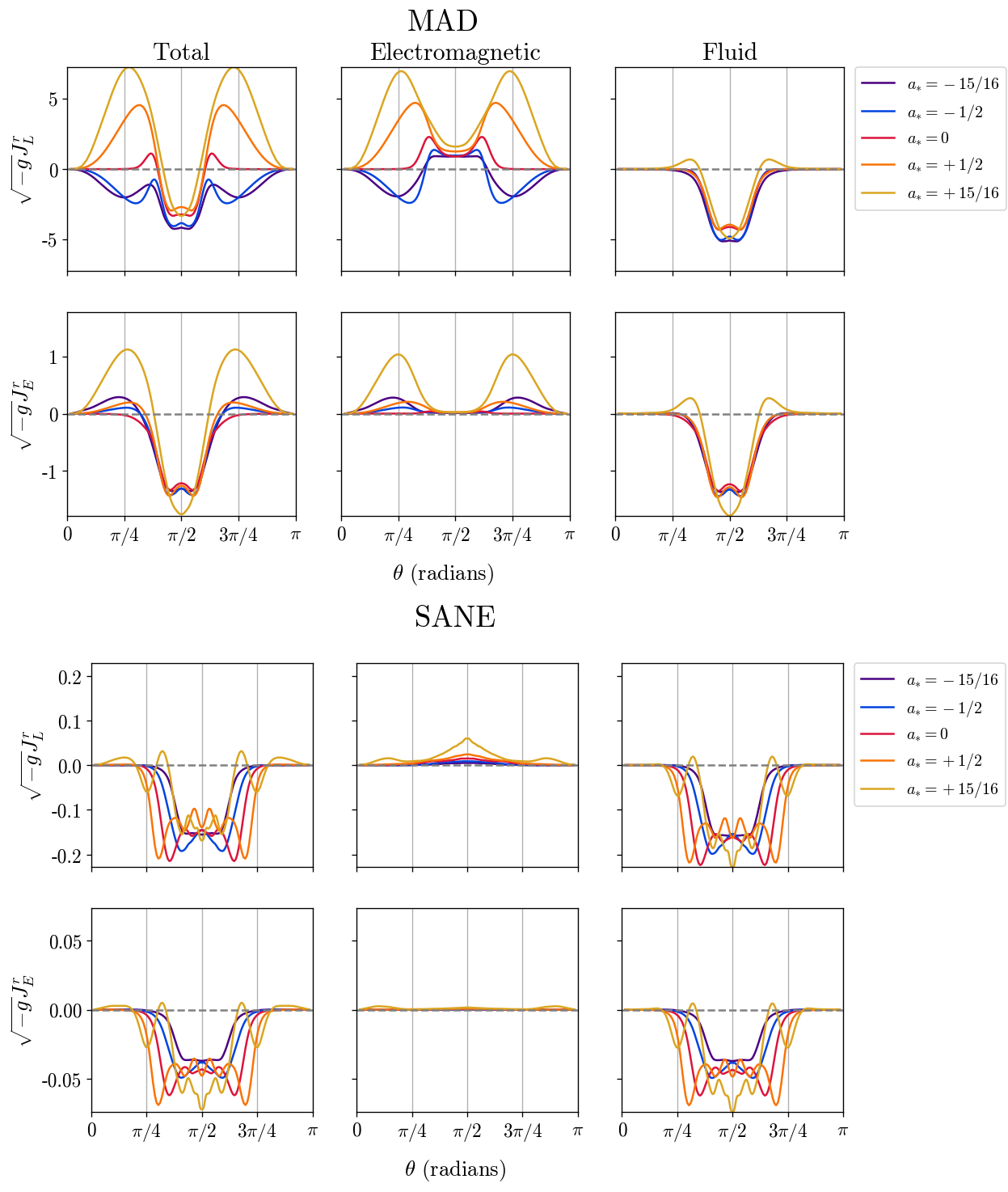}
\caption{Similar to Figure \ref{fig:conserved_currents_colatitude_radial_dependence}, but instead we plot the conserved current profiles at a fixed radius $r=10~r_{g}$ for all simulations in the library. The outward transport of angular momentum and energy in the jet strongly depends on the black hole spin for the MAD models. Negative values of $\sqrt{-g}J^{r}_{L}$ at higher latitudes for the retrograde models is simply a reflection of the magnetic fieldlines and plasma in the jet corotating with the black hole.}
\label{fig:conserved_currents_colatitude}
\end{figure}

The motion of ideal fluids in a stationary and axisymmetric metric gives rise to three \textit{conserved currents}. Particle number conservation generates a mass current $J^{\mu}_{M} \equiv \rho u^{\mu}$ which is conserved along the flow. An equivalent coordinate-dependent current density $\boldsymbol{\mathfrak{J}}_{M} \equiv \sqrt{-g}\boldsymbol{J}_{M}$ can also be defined, which will prove beneficial as we plot and compare quantities in a coordinate frame. The Kerr metric is independent of $t$ and $\phi$, i.e., there exist one-parameter family of diffeomorphisms or ``isometries'' generated by associated Killing vector fields  $\boldsymbol{\xi}_{t}$ and $\boldsymbol{\xi}_{\phi}$ such that the Lie derivative of the metric along these vector fields is zero. In the Kerr-Schild coordinates these vector fields are given by,
\begin{equation}
    \xi^{\mu}_{t} = \{1,0,0,0\},\hspace{0.1in}\xi^{\mu}_{\phi} = \{0,0,0,1\}.
\end{equation}
The Killing equation $\nabla_{(\mu}\xi_{\nu)}=0$ combined with the conservation of the stress-energy tensor gives rise to two additional currents: $J^{\mu}_{E}\equiv-\xi^{\nu}_{t}T^{\mu}_{\nu}$, representing the flow of total energy, and $J^{\mu}_{L}\equiv\xi^{\nu}_{\phi}T^{\mu}_{\nu}$, representing the flow of angular momentum. The corresponding current densities are $\boldsymbol{\mathfrak{J}}_{E} \equiv \sqrt{-g}\boldsymbol{J}_{E}$ and $\boldsymbol{\mathfrak{J}}_{L} \equiv \sqrt{-g}\boldsymbol{J}_{L}$. Henceforth, we will refer to these current densities simply as currents, as we focus on coordinate-dependent quantities.

In this section we compute conserved currents for the \v3 simulation library and examine trends in the flow of mass, angular momentum and energy as a function of black hole spin and magnetization state. The first row in Figure \ref{fig:conserved_currents_mad+0.94} plots the time- and azimuth-averaged conserved currents for the MAD $a_{*}=+0.94$ model. Streamlines are overlaid on the average rest-mass density to help interpret the structure of the conserved currents in relation to the overall flow. The colorscale of the streamlines indicates the magnitude of the conserved currents, spaced logarithmically. Advection-dominated accretion drives strong inflow of mass, angular momentum and energy in the disk. Moving to higher latitudes, the flow lines shift from ingoing to outgoing, with the reversal in the mass current direction indicating the presence of weak MHD winds that produce modest mass outflows. The outflowing streamlines of angular momentum and energy currents, originating from the inner accretion flow region ($r\lesssim10$ $r_{g}$) are highly collimated and anchored at the black hole horizon. Maximal outflow follows a parabolic profile, suggesting the winding of magnetic field lines around the polar axis through the Blandford-Znajek (BZ) mechanism \citep{blandford_electromagnetic_1977}.

To assess the electromagnetic and fluid contributions to the conserved currents, we perform the following decomposition,
\begin{align} \label{eqn:conserved_current_L_em_fluid_decomposition}
\begin{split}
    J^{\mu}_{\mathrm{L,EM}} &=  \langle b^{2}u^{\mu}u_{\phi} - b^{\mu}b_{\phi} + \delta^{\mu}_{\phi}b^{2}/2\rangle, \\
    J^{\mu}_{\mathrm{L,Fl}} &=  \langle (\rho + u + p_{g})u^{\mu}u_{\phi} + \delta^{\mu}_{\phi}p_g\rangle,
\end{split}
\end{align}
and,
\begin{align}\label{eqn:conserved_current_E_em_fluid_decomposition}
\begin{split}
    J^{\mu}_{\mathrm{E,EM}} &=  \langle -(b^{2}u^{\mu}u_{t} - b^{\mu}b_{t} + \delta^{\mu}_{t}b^{2}/2)\rangle, \\
    J^{\mu}_{\mathrm{E,Fl}} &= \langle -(\rho + u + p_{g})u^{\mu}u_{t} - \delta^{\mu}_{t}p_g\rangle,
\end{split}
\end{align}
where $\delta^{\mu}_{\nu}$ is the Kronecker delta and $\langle\dots\rangle$ denotes an average over $t$ and $\phi$. Unlike Equation \ref{eqn:density_weighted_radial_profiles}, we do not consider a density-weighted average here. Previous studies that investigate the transport of angular momentum in global accretion flows further breakdown the electromagnetic and fluid contributions of $J^{\mu}_{L}$ into a laminar and turbulent component \citep{bethune_protoplanetary_2017, mishra_mad_2020, jacquemin-ide_magnetic_outflows_2021, manikantan_mad_transport_2023,scepi_thin_mad_2024, jacquemin-ide_mri_2024} or an advective and stress component \citep{penna_thin_disc_2010, chatterjee_angular_momentum_2022}. In this study we restrict our analysis to the \textit{total} electromagnetic and fluid portion as defined in Equation \ref{eqn:conserved_current_L_em_fluid_decomposition}.


The partitioning of $\boldsymbol{\mathfrak{J}}_{L}$ and $\boldsymbol{\mathfrak{J}}_{E}$, as defined in Equations \ref{eqn:conserved_current_L_em_fluid_decomposition} and \ref{eqn:conserved_current_E_em_fluid_decomposition}, is plotted in Figure \ref{fig:conserved_currents_mad+0.94} for MAD $a_{*}=+0.94$. Significant efflux of electromagnetic angular momentum and energy in the jet, following the parabolic profile shown in the total currents respectively, supports BZ mechanism. This is seen clearly in the top two rows of Figure \ref{fig:conserved_currents_colatitude_radial_dependence} where we plot the radial components of the currents for the same model as a function of colatitude $\theta$. The shift in the peak toward higher latitudes with increasing radius highlights the collimated structure of the jet.

Figure \ref{fig:conserved_currents_sane+0.5} is identical to Figure \ref{fig:conserved_currents_mad+0.94} but for SANE $a_{*}=+0.5$. We see that the conserved currents are directed radially inward across all latitudes implying the absence of any significant outflow \footnote{While there is some outflow in $\boldsymbol{\mathfrak{J}}_{L}$ close to the poles, the colorscale, which plots the magnitude of poloidal components of the current, suggests this is negligible (recall the colorscale is logarithmically spaced).}. The fluid component of the angular momentum and energy currents emphasizes the advective nature of the flow, while magnetic stresses facilitate outward transport of angular momentum in the disk through the MRI. The colatitude plots in Figure \ref{fig:conserved_currents_colatitude_radial_dependence} underscores this point.

Finally, in Figure \ref{fig:conserved_currents_colatitude} we compare $\boldsymbol{\mathfrak{J}}_{L}$ and $\boldsymbol{\mathfrak{J}}_{E}$ across all simulations in our library by plotting $\mathfrak{J}^{r}(\theta)$ at $r=10~r_{g}$. In the MAD models, the outward transport of angular momentum and Poynting flux in the magnetically dominated jet shows a strong dependence on black hole spin. Note that negative values for $\mathfrak{J}^{r}$ in the jet for retrograde models result from the corotation of the magnetic field and plasma with the black hole, i.e., $u_{\phi}<0$ (see \citealt{wong_jet-disk_2021}). $\mathfrak{J}^{r} (\theta)$ shows a weak dependence on black hole spin for SANE models.
\subsection{Jet power}
\label{sec:jet power}

Global simulations of black hole accretion produce relativistic outflows of highly magnetized material ``jets'' along the black hole spin axis (see e.g., \citealp{mckinney_measurement_2004, tchekhovskoy_efficient_2011}; also see \citealp{davis_review_jets_2020, komissarov_jets_2021} for a recent review of jets in numerical simulations) \footnote{The picture is less clear for tilted accretion flows, where the angular momentum vector of disk is not necessarily aligned with that of the black hole. \cite{liska_tilted_grmhd_2018, chatterjee_tilted_images_2020} find the jet aligns with the disk's angular momentum vector, while \cite{ressler_windfed_tilt_2023} observe the jet aligned with the black hole spin axis in the case of MAD accretion flows.}. These jets are consistent with the Blandford-Znajek picture where the field lines that are brought in by the accreting plasma are dragged by the black hole in its ergosphere before eventually being launched carrying energy and momentum outward.

We follow \citetalias{M87PaperV} and define the jet as the region (within 1 radian of the pole) that satisfies,
\begin{equation}
\label{eqn:betagamma_cut}
(\beta\gamma)^2\equiv\bigg(\frac{-T^{r}_{t}}{\rho u^{r}}\bigg)^2 - 1 > (\beta\gamma)_{\text{cut}}^2,
\end{equation}
where $-T^r_t$ and $\rho u^r$ are time- and azimuth-averaged. $(\beta\gamma)_{\text{cut}}$ is set to unity but the choice of the cut is arbitrary and Figure 10 in \citetalias{M87PaperV} explores the effect of changing this value. The total mechanical power in the relativistic jet is then defined as,
\begin{equation}
\label{eqn:jet_power}
    P_{\mathrm{jet}} = \int_{\beta\gamma>(\beta\gamma)_{cut}}\mathrm{d}\theta\frac{1}{\Delta t}\int \mathrm{d}t\mathrm{d}\phi\sqrt{-g}(-T^r_t-\rho u^r).
\end{equation}
measured at $r=100~r_g$. Additionally, we also define a total outflow power $P_{\text{out}}$ that is computed in the same manner as Equation \ref{eqn:jet_power} but without the $(\beta\gamma)_{\text{cut}}$. The $\theta$ integral is instead carried out over 1 radian about the poles. This diagnostic includes both the fast-moving collimated jet, and the slow-moving winds further away from the poles. There are several ways of defining a jet, or more broadly, an unbound outflow, and we discuss this further in Appendix \ref{appendix:jet_power}.

We report jet and outflow power for the \v3 library in Table \ref{table:jet_outflow_power_statistics}. We see that according to our definition of $P_{\text{jet}}$, low spinning SANEs do not produce an appreciable jet, but do contain slow outgoing winds that result in a non-zero $P_{\text{out}}$. MAD $a_{*}=+0.94$ has the most powerful jet, an outcome that is expected if the driving mechanism is the Blandford-Znajek process (where $P_{\text{jet}}\propto a_{*}^{2}\Phi_{\mathrm{BH}}^{2}$; see e.g. \citealt{mckinney_bz_power_2005,tchekhovskoy_dichotomy_agn_2010}). Finally, we note that for models where $P_{\text{jet}}\neq0$, this quantity is dominated by the Poynting flux.

\setlength{\tabcolsep}{8pt}
\begin{deluxetable}{ lrcccc }
\tablecaption{Jet power statistics} \label{table:jet_outflow_power_statistics}
\tablehead{
\colhead{Flux} & 
\colhead{$a_*$} &
\colhead{$P_{\text{jet}}/(\dot{M}c^2)$} &
\colhead{$P_{\text{out}}/(\dot{M}c^2)$} &
\colhead{$P_{\text{jet,em}}/P_{\text{jet}}$}
}
\startdata
MAD & \(-15/16\) & \(3.50\times10^{-1}\) & \(3.83\times10^{-1}\) & \(7.22\times10^{-1}\) \\ 
MAD & \(-1/2\) & \(9.08\times10^{-2}\) & \(1.27\times10^{-1}\) & \(7.83\times10^{-1}\) \\ 
MAD & 0 & \(6.02\times10^{-5}\) & \(2.27\times10^{-2}\) & \(6.10\times10^{-1}\) \\ 
MAD & \(+1/2\) & \(1.97\times10^{-1}\) & \(2.49\times10^{-1}\) & \(8.18\times10^{-1}\) \\ 
MAD & \(+15/16\) & 1.51 & 1.60 & \(7.11\times10^{-1}\) \\ 
SANE & \(-15/16\) & \(2.15\times10^{-3}\) & \(2.46\times10^{-3}\) & \(7.64\times10^{-1}\) \\ 
SANE & \(-1/2\) & 0.00 & \(3.53\times10^{-4}\) & - \\ 
SANE & 0 & 0.00 & \(3.96\times10^{-4}\) & - \\ 
SANE & \(+1/2\) & 0.00 & \(8.37\times10^{-4}\) & - \\ 
SANE & \(+15/16\) & \(3.30\times10^{-2}\) & \(4.32\times10^{-2}\) & \(6.45\times10^{-1}\) \\
\enddata
\tablecomments{Summary of jet and outflow power. The first two columns are identical to Table \ref{table:grmhd_models} where Flux labels the strength of the magnetic field at the horizon and $a_{*}$ is black hole spin. The third and fourth column are jet and outflow efficiencies respectively, and the final column is the electromagnetic fraction of the jet power.}
\end{deluxetable}

\subsection{Black hole spin-up/spin-down}
\label{sec:spinup}

\begin{figure}
\centering
\includegraphics[,width=\linewidth]{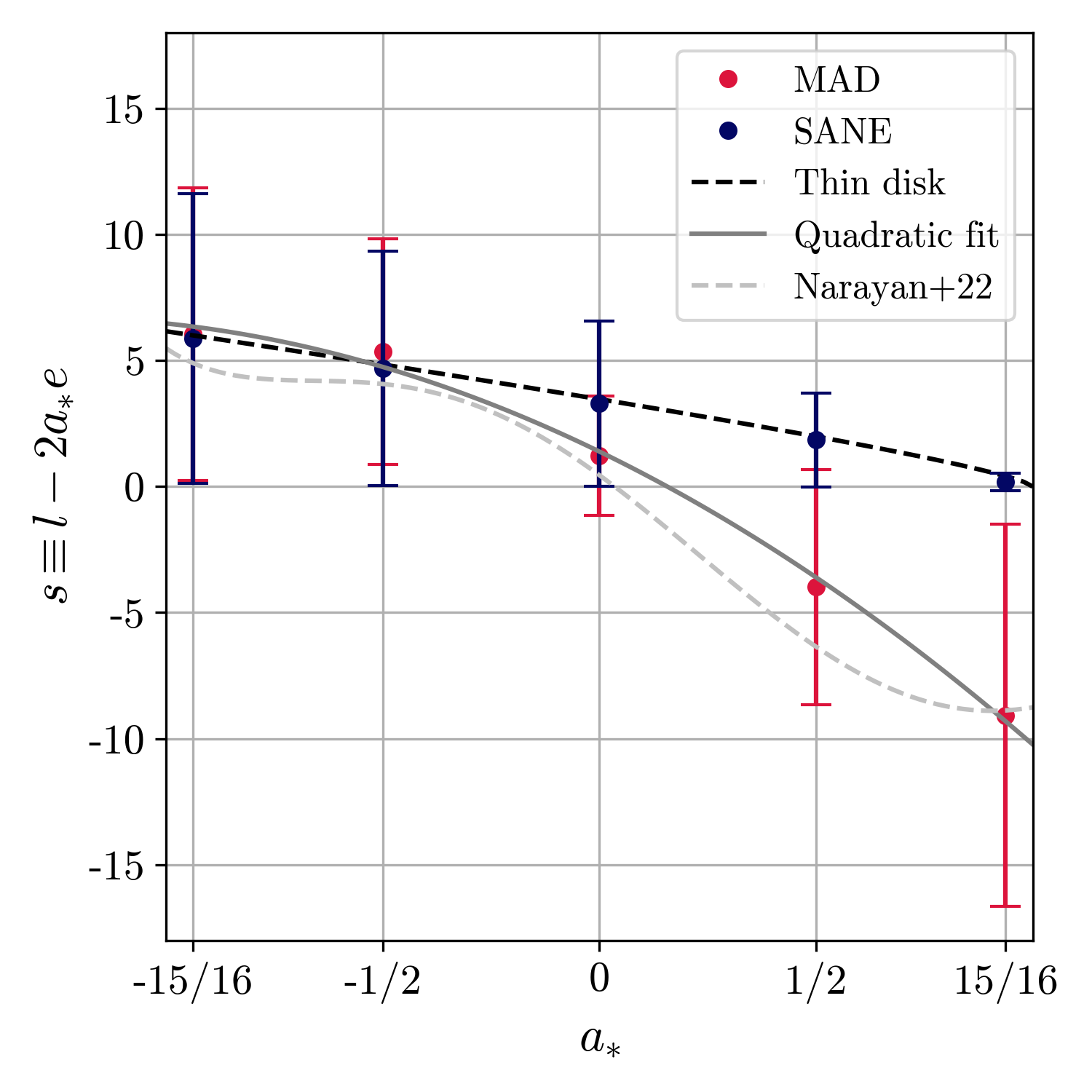}
\caption{The spin-up parameter `$s$' given by Equation \ref{eqn:spinup_defintion}. The markers indicate the time-averaged value and error bars represent 1$\sigma$. Red and blue markers are MAD and SANE simulations respectively. The black dashed line is the expected thin disk values (Equation \ref{eqn:spinup_thin_disk}). The solid gray line is the quadratic fit (Equation \ref{eqn:spinup_mad_fit}) and the dashed silver line is the fifth-order polynomial fit provided in \cite{narayan_jets_2022}.}
\label{fig:spinup}
\end{figure}

\cite{bardeen_spinup_1970} was the first to consider the role of the surrounding environment on the properties of the central black hole. He assumed a cold equatorial accretion disk with matter plunging from the ISCO onto the black hole. He showed that this leads to the black hole achieving the maximal spin $a_{*}=1$ in finite time. \cite{thorne_spinup_1974} included the effect of radiation emitted from the disk on the spin-up of the black hole. The photons captured by the black hole provide a counteracting torque that allows a maximal spin-up of $a_{*}\simeq0.998$. The author highlights the assumptions in their model, particularly regarding the potential sensitivity of the results to the disk thickness ($h/r$) and the presence of local magnetic fields.

\cite{gammie_spinup_2004} used GRMHD simulations to address the question of black hole spin-up. They considered axisymmetric SANE simulations to identify spin equilibrium (defined as $s=0$; see also \citealt{shapiro_spinup_2005}) where,
\begin{equation}
    s\equiv\frac{da_{*}}{dt}\frac{M}{\dot{M}}.
\end{equation}
This may be expressed in terms of the ingoing specific angular momentum and energy flux as,
\begin{equation}
\label{eqn:spinup_defintion}
    s=l-2a_{*}e.
\end{equation}
They find spin equilibrium at $a_{*}\simeq0.94$. \cite{tchekhovskoy_general_2012} and more recently \cite{narayan_jets_2022, lowell_spindown_2023} found the spin equilibrium in MAD simulations is significantly lower, with $a_{\text{eq}}\lesssim0.1$. This is attributed to strong magnetic fields in the jet, which extract angular momentum from the black hole more efficiently than that supplied by disk.

In Figure \ref{fig:spinup} we plot the time-averaged spin-up parameter. We see that the SANE simulations, where the magnetic fields are dynamically unimportant and primarily serve to transport angular momentum outward through the disk, closely follow the thin disk expectation,
\begin{equation}
\label{eqn:spinup_thin_disk}
    s_{\text{thin}} = l_{\text{isco}} - 2a_{*}e_{\text{isco}}.
\end{equation}
However, the role of magnetic fields is not ignorable as the models achieve spin equilibrium at $a_{\text{eq}}\sim0.94$, confirming earlier work. Prograde MAD models possess a negative spin-up parameter indicating black hole spin-down. The spin-down is stronger for more rapidly spinning black holes due to the presence of stronger radial magnetic fields anchored at the event horizon. We provide a quadratic fit to our models,
\begin{equation}
\label{eqn:spinup_mad_fit}
    s_{\text{fit}} = -3.267a_{*}^2 - 8.349a_{*} + 1.387,
\end{equation}
and also plot the fifth-order fit from \cite{narayan_jets_2022} \footnote{Note that our sampling of the black hole spin space is sparser than \cite{narayan_jets_2022}, and we consider a lower order polynomial for our fit.}. The retrograde models spin up the black hole; however, since $a_{*}<0$ for these models, the absolute value of spin decreases. Finally, we observe that the time variation in $s$ (indicated by the errorbars)  is more pronounced for MADs, likely due to the intermittent nature of accretion flow in these simulations.
\section{Simulation data products}
\label{sec:data_products}

The \v3 GRMHD simulations are staged on a local file server at the University of Illinois at Urbana-Champaign. Interested readers may access the data using the following URL: \url{http://thz.astro.illinois.edu/}. A Python package for analyzing the GRMHD data is publicly available \footnote{\url{https://github.com/AFD-Illinois/pyharm/}}. Due to constraints on storage space, we host simulation snapshots corresponding to $25,000-30,000\hspace{0.1cm}t_{g}$, but the full dataset will be shared upon reasonable request to the corresponding author. 
\section{Summary}
\label{sec:summary}

In this paper we present data products from a suite of ideal GRMHD simulations of black hole accretion. The library consists of 10 simulations that cover the accretion disk magnetization state `$\phi_{b}$', black hole spin `$a_{*}$' parameter space. We consider 5 spins $a_{*} = 0,\pm0.5,\pm0.94$ for each state of magnetization (MAD and SANE). The GRMHD simulations discussed in this paper were used in the Galactic center analysis by the EHT (\citetalias{SgrAPaperV,SgrAPaperVIII}) and in other works \citep{georgiev_variability_2022,conroy_pattern_speed_2023,chan_variability_2024,chan_variability_origin_2024,joshi_cp_2024}. We studied trends across the simulations and summarize our findings below:
\begin{enumerate}
    \item The horizon-penetrating, normalized radial angular momentum flux in SANE simulations closely matches the thin-disk expectation. In prograde MAD simulations, magnetic field lines anchored at the horizon extract angular momentum from the black hole, and this overwhelms the inflow of positive angular momentum from the disk. This result is consistent with the findings of the ``spin-up'' study, where prograde MAD models experience spin-down, i.e., a reduction in the black hole’s angular momentum. SANE models follow the thin-disk expectation closely and achieve spin equilibrium at $a_{*}\approx0.94$.
    \item For $a_{*}\gtrsim 0$, the thermodynamic component of the specific energy flux is no longer negligible in the SANE simulations, and the total specific energy flux for prograde models exceeds the thin disk expectation. This is further evidenced by Figure \ref{fig:disk_avg_radial_profiles}, which shows an increase in fluid temperature near the black hole with increasing spin. The $a_{*}=+0.94$ MAD model has an outflow efficiency $>1$,  indicating that the energy outflow in jets exceeds the energy supplied by the accretion disk.
    \item Analysis of the time series of radial fluxes at the horizon reveals that MAD simulations exhibit greater variability on average than their SANE counterparts. This aligns with the physical picture in which MAD disks undergo aperiodic flux eruption events, characterized by a drop in $\Phi_{\mathrm{BH}}$ and a subsequent rise in $\dot{M}$, whereas SANE accretion is more uniform. For both SANE and MAD simulations we find that the average $\dot{M}$ variability increases with spin for prograde models, but is independent of spin for retrograde models, consistent with the findings of \cite{narayan_jets_2022}.
    \item A time- and azimuth-averaged study of the accretion disk shows that MAD disks are consistently hotter than SANE disks within $r\lesssim10\hspace{0.1cm}r_{g}$. Prograde MAD models produce more powerful and wider jets than their retrograde counterparts, resulting in narrower disks for retrograde models. We find the spin-dependence of temperature, specific angular momentum $u_{\phi}$, and disk scale height $h/r$ in SANE simulations closely aligns with the semi-analytic model of relativistic, advection-dominated viscous accretion flow provided by \cite{popham_gammie_adaf_1998}. Additionally, MAD disks are sub-Keplerian with significantly larger variance in angular velocity as compared to SANE disks.
    \item We examine the transport of mass, angular momentum, and energy in our simulations by analyzing conserved currents $J_{M}=\rho u^{\mu}$, $J_{L}=T^{\mu}_{\phi}$, and $J_{E}=-T^{\mu}_{t}$ respectively. In MAD simulations, dynamically significant magnetic fields facilitate the outward transport of angular momentum and energy through the jet and winds. Within the disk, the fluid sector primarily drives the inward transport of angular momentum and energy. In SANE simulations, the electromagnetic component contributes to outward angular momentum transport within the disk, indicating the influence of MRI. 
\end{enumerate}

Finally, it is worth providing an honest assessment of the limitations of this work. This is in addition to the caveats associated with ideal GRMHD discussed in Section \ref{subsec:grmhd_asusmptions_caveats}. The spatial resolution of the simulations in the \v3 library is somewhat lower than contemporary studies (see, e.g., \citet{dexter_parameter_2020,mizuno_comparison_2021,narayan_jets_2022,fromm_nonthermal_2022}). Resolution studies indicate that while time-averaged, bulk properties of MAD accretion flows are converged at this resolution \citep{white_mad_resolution_study_2019,salas_mad_resolution_study_2024}, the spatial structure of the flow at higher latitudes, such as the jet-disk boundary, is not converged. This can be important when interpreting VLBI observations of edge-brightened jets \citep{walker_edge_brightened_jet_2018,kim_limb_brightened_jet_2018,lu_ring_jet_2023,davelaar_synchrotron_waves_boundary_2023,kim_limb_brightened_jet_2024}. In SANE simulations, where the MRI facilitates outward transport of angular momentum, \cite{shiokawa_global_2012,porth_cc_2019} find that the $\alpha$-parameter does not converge over the range of resolutions considered. Additionally, \cite{ripperda_flares_2022} find higher-resolution simulations necessary to achieve convergence in the magnetic reconnection rate.

To minimize differences in synthetic observables arising from code-dependent choices, we chose $\hat{\gamma}=4/3$ to be consistent with other GRMHD codes considered in \citetalias{SgrAPaperV}. The lack of an efficient coupling mechanism between the ions and electrons in RIAFs results in a two-temperature plasma with relativistic electrons and nonrelativistic ions. This suggests that in single-temperature simulations, the fluid adiabatic index should be close to 5/3 (C. F. Gammie 2024, in prep). 

Although this library evolves simulations for a significantly extended period compared to previous studies used in EHT analysis \citep{Wong_2022_patoka}, even longer runs will be necessary to achieve inflow equilibrium at larger radii. This may be crucial as Faraday rotation effects arising from outer regions of the accretion disk can undermine polarimetric images. Finally, a finer sampling of black hole spin, i.e., an increased number of GRMHD simulations with different $a_{*}$, may provide an independent constraint on black hole spin (V. Bernshteyn et al. 2024, in prep).

\acknowledgements
\label{sec:Acknowledgements}
V.D. is grateful to Abhishek Joshi for discussions that greatly improved the quality of certain sections in this text. V.D. was supported in part by the ICASU/NCSA Fellowship. G.N.W.~was supported by the Taplin Fellowship and the Princeton Gravity Initiative.
This work was supported by NSF grants AST 17-16327 (horizon), OISE 17-43747, and AST 20-34306. This research used resources of the Oak Ridge Leadership Computing Facility at the Oak Ridge National Laboratory, which is supported by the Office of Science of the U.S. Department of Energy under Contract No. DE-AC05-00OR22725. This research used resources of the Argonne Leadership Computing Facility, which is a DOE Office of Science User Facility supported under Contract DE-AC02-06CH11357. This research is part of the Delta research computing project, which is supported by the National Science Foundation (award OCI 2005572), and the State of Illinois. Delta is a joint effort of the University of Illinois at Urbana-Champaign and its National Center for Supercomputing Applications. The data analysis was possible thanks to the high throughput computing utility `Launcher' \citep{wilson_launcher_2017}. This work used the Extreme Science and Engineering Discovery Environment (XSEDE), which is supported by National Science Foundation grant number ACI-1548562, specifically the XSEDE resources Longhorn, Frontera, and Stampede2 at the Texas Advanced Computing Center (TACC) through allocation TG-AST170024. The authors acknowledge the Texas Advanced Computing Center (TACC) at The University of Texas at Austin for providing computational resources that have contributed to the research results reported within this paper.

\appendix

\section{Calculating the magnetic flux threading the black hole horizon}\label{appendix:phi_bh}

\begin{figure*}
\centering
\includegraphics[,width=\linewidth]{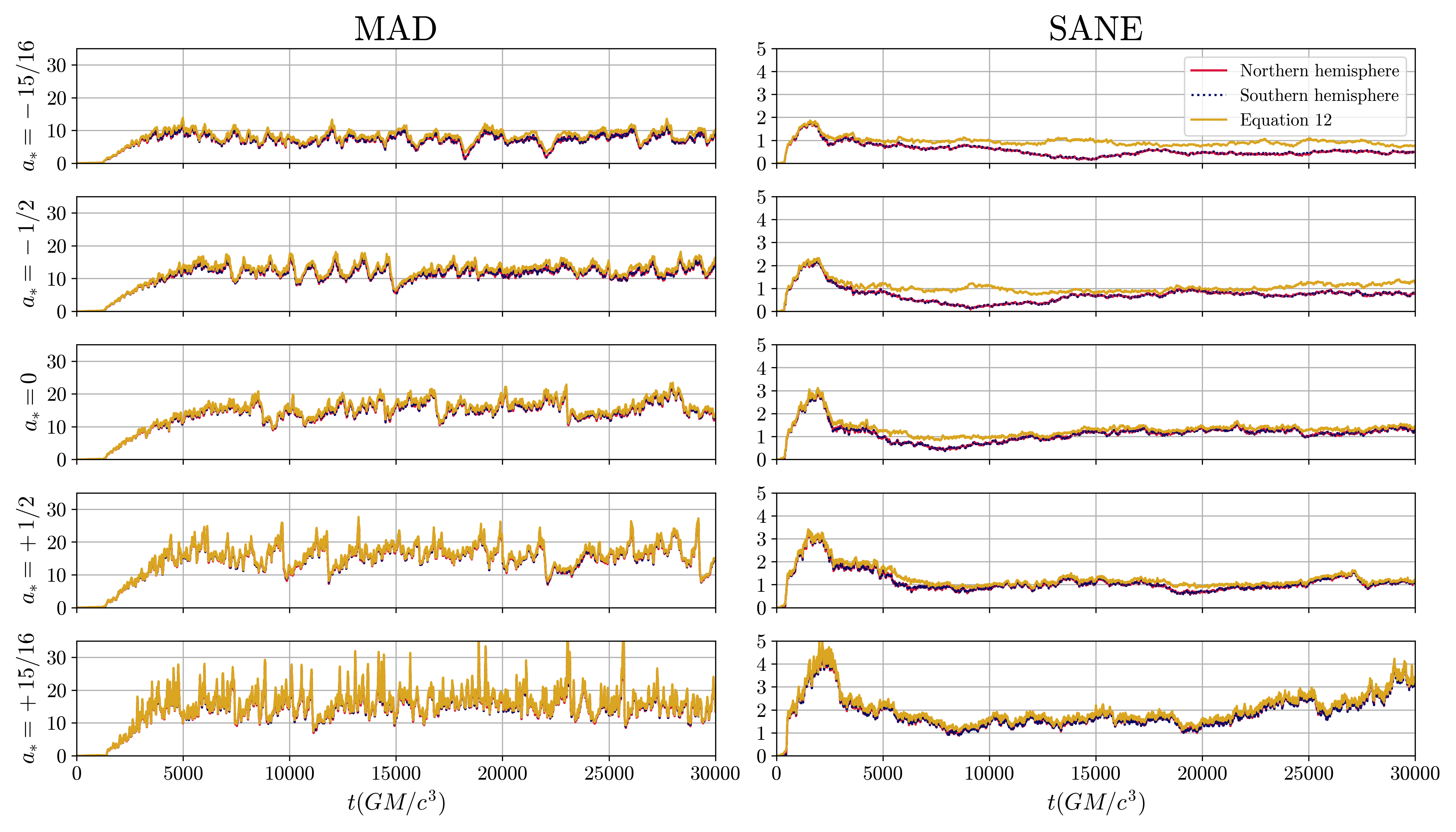}
\caption{A time series of the dimensionless magnetic flux crossing the event horizon for all the models in \texttt{v3}. Columns: Left: MAD, Right: SANE simulations. Rows: Top to bottom: Increasing spin from $a_{*}=-15/16$ to $a_{*}=+15/16$. The gray line plots the flux as computed according to Equation \ref{eqn: magnetic_flux_definition}, while the goldenrod and magenta lines plot $\Phi_{\text{BH}}^{\text{N}}$ (Equation \ref{eqn: magnetic_flux_definition_north}) and $\Phi_{\text{BH}}^{\text{S}}$ (Equation \ref{eqn: magnetic_flux_definition_south}) respectively.}
\label{fig:phibh_time_series}
\end{figure*}

The standard approach for calculating the magnetic flux at the event horizon is given by Equation \ref{eqn: magnetic_flux_definition}. This approach assumes that the total flux crossing each hemisphere of the horizon---north ($\theta < \pi/2$) and south ($\theta\geq\pi/2$)---is equal in magnitude and opposite in direction. However, this assumption may not hold if the magnetic field changes direction in certain regions within either hemisphere.

To investigate this, we compute the magnetic flux across the northern and southern hemispheres using the following definitions:
\begin{align}
    \Phi_{\text{BH}}^{\text{N}} &= \int_{\phi}\int_{\theta<\pi/2} B^r\sqrt{-g}\:d\theta\:d\phi, \label{eqn: magnetic_flux_definition_north}\\
    \Phi_{\text{BH}}^{\text{S}} &=  \int_{\phi}\int_{\theta\geq\pi/2} -B^r\sqrt{-g}\:d\theta\:d\phi, \label{eqn: magnetic_flux_definition_south}
\end{align}
and compare them with the standard definition. The negative sign in Equation \ref{eqn: magnetic_flux_definition_south} ensures that $\Phi_{\text{BH}}^{\text{N}}$ and $\Phi_{\text{BH}}^{\text{S}}$ are consistent.

In Figure \ref{fig:phibh_time_series} we plot $\phi_{b}$ using Equations \ref{eqn: magnetic_flux_definition}, \ref{eqn: magnetic_flux_definition_north}, and \ref{eqn: magnetic_flux_definition_south}. The flux across both the hemispheres (Equations \ref{eqn: magnetic_flux_definition_north} and \ref{eqn: magnetic_flux_definition_south}) agrees closely for all the models, with relative differences being less than $10^{-6}$. The standard definition of the flux emulates the hemispherical values for all MAD simulations and prograde SANE models. However, for $a_{*}\leq 0$ SANE models this is not the case, and at several points in the simulation the standard calculation overestimates the individual hemispherical flux estimates.

\begin{figure}
\centering
\includegraphics[,width=\linewidth]{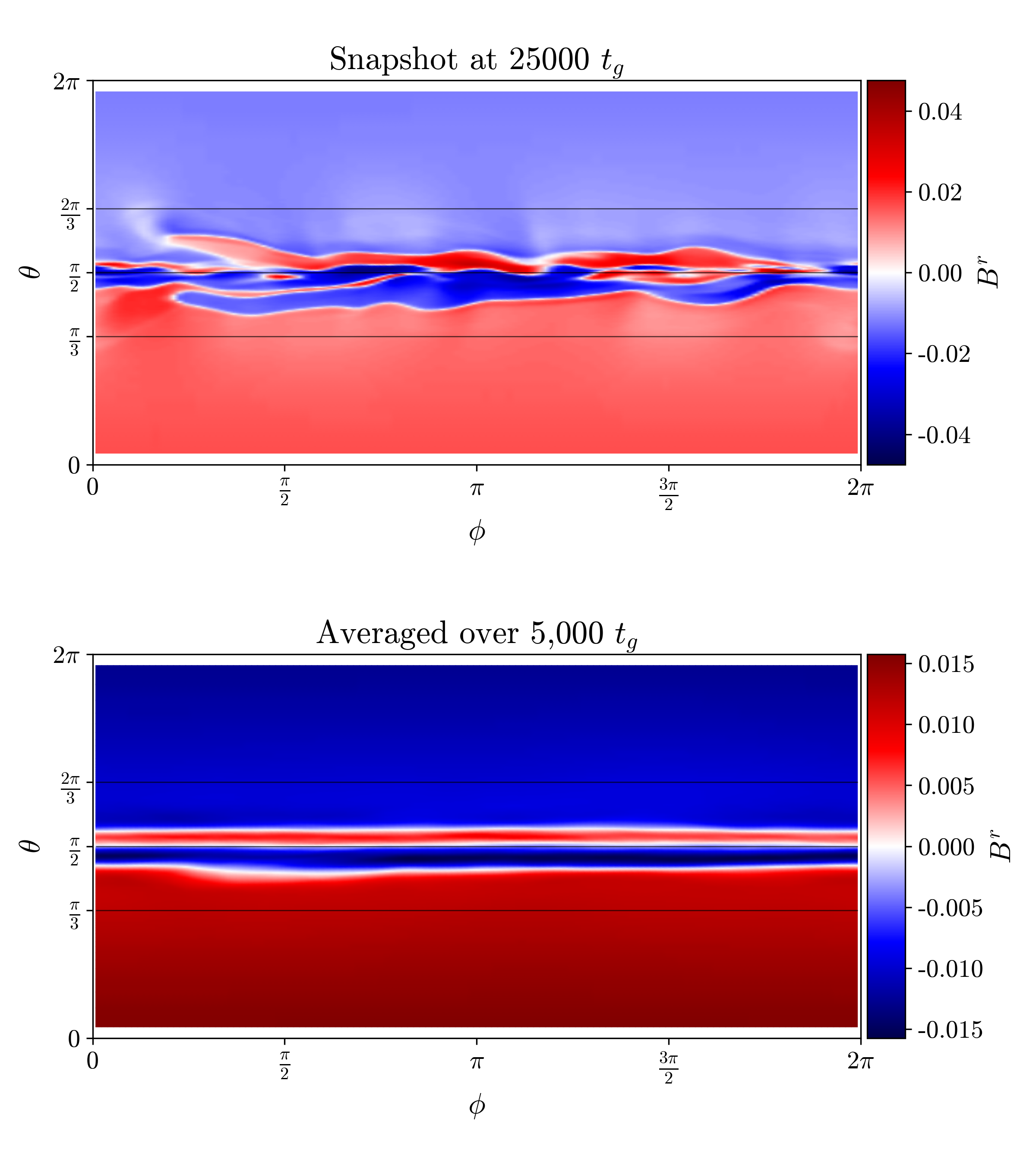}
\caption{A heat map of the radial component of the magnetic field $B^{r}$ at the event horizon for the SANE $a_{*}=-0.5$ simulation. Top panel: $B^{r}$ at a specific instant in time. Bottom panel: $B^{r}$ averaged over the interval $t=[25, 30] \times 10^3 t_{g}$. In both panels, note that $B^{r}$ changes polarity within each hemisphere. This behavior is not accounted for in Equation \ref{eqn: magnetic_flux_definition} leading to an overestimation of $\Phi_{\text{BH}}$.}
\label{fig:br_horizon_sane-0.5}
\end{figure}

\begin{figure}
\centering
\includegraphics[,width=\linewidth]{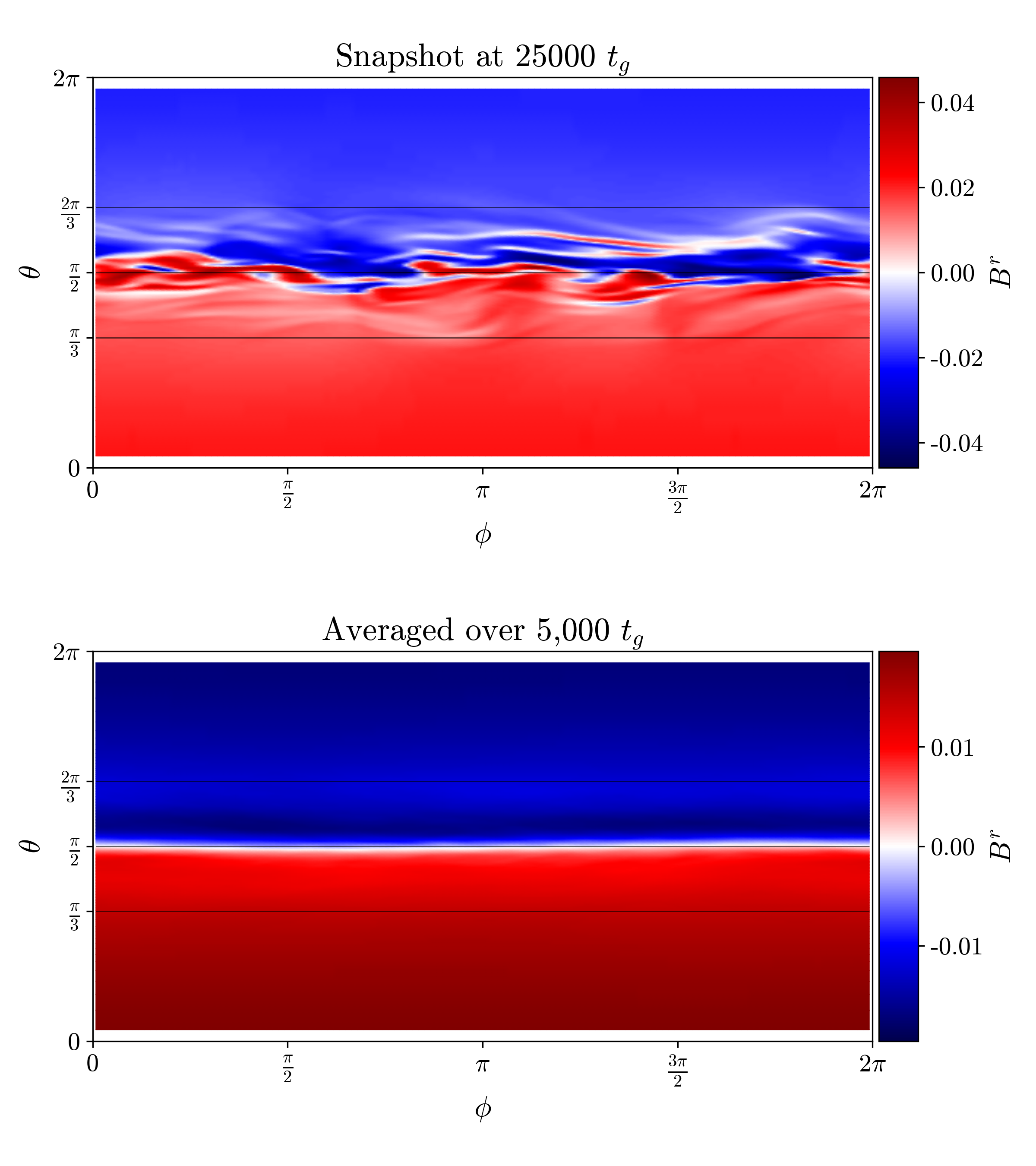}
\caption{Same as Figure \ref{fig:br_horizon_sane-0.5} but for SANE $a_{*}=+0.5$. The magnetic fields in this case are much more ordered, resulting in a consistent sign of $B^{r}$ across both hemispheres.}
\label{fig:br_horizon_sane+0.5}
\end{figure}

To better understand this, we plot the radial component of the magnetic field $B^{r}$ for two models: one where the values do not match, SANE $a_{*}=-0.5$ (Figure \ref{fig:br_horizon_sane-0.5}), and one where they do, SANE $a_{*}=+0.5$  (Figure \ref{fig:br_horizon_sane+0.5}). In the retrograde model we see a quadrupolar structure in $B^{r}$ that persists over time. This polarity reversal within each hemisphere explains the reduced magnetic flux when computed using Equations \ref{eqn: magnetic_flux_definition_north} and \ref{eqn: magnetic_flux_definition_south} which accounts for the sign reversal. In contrast, for the prograde model, the time-averaged $B^{r}$ maintains a constant sign in each hemisphere \footnote{We observe a numerical artifact near the poles associated with meshblock boundaries in one of the simulations (SANE $a_{*}=-0.94$). This issue has been resolved in \kharma since the \v3 library was generated. We reran this simulation and confirmed that the artifact does not impact the bulk properties of the disk. The new simulation is also available at \url{http://thz.astro.illinois.edu/}.}.

\section{Jet Power Definition}\label{appendix:jet_power}

\begin{figure}
\centering
\includegraphics[,width=\linewidth]{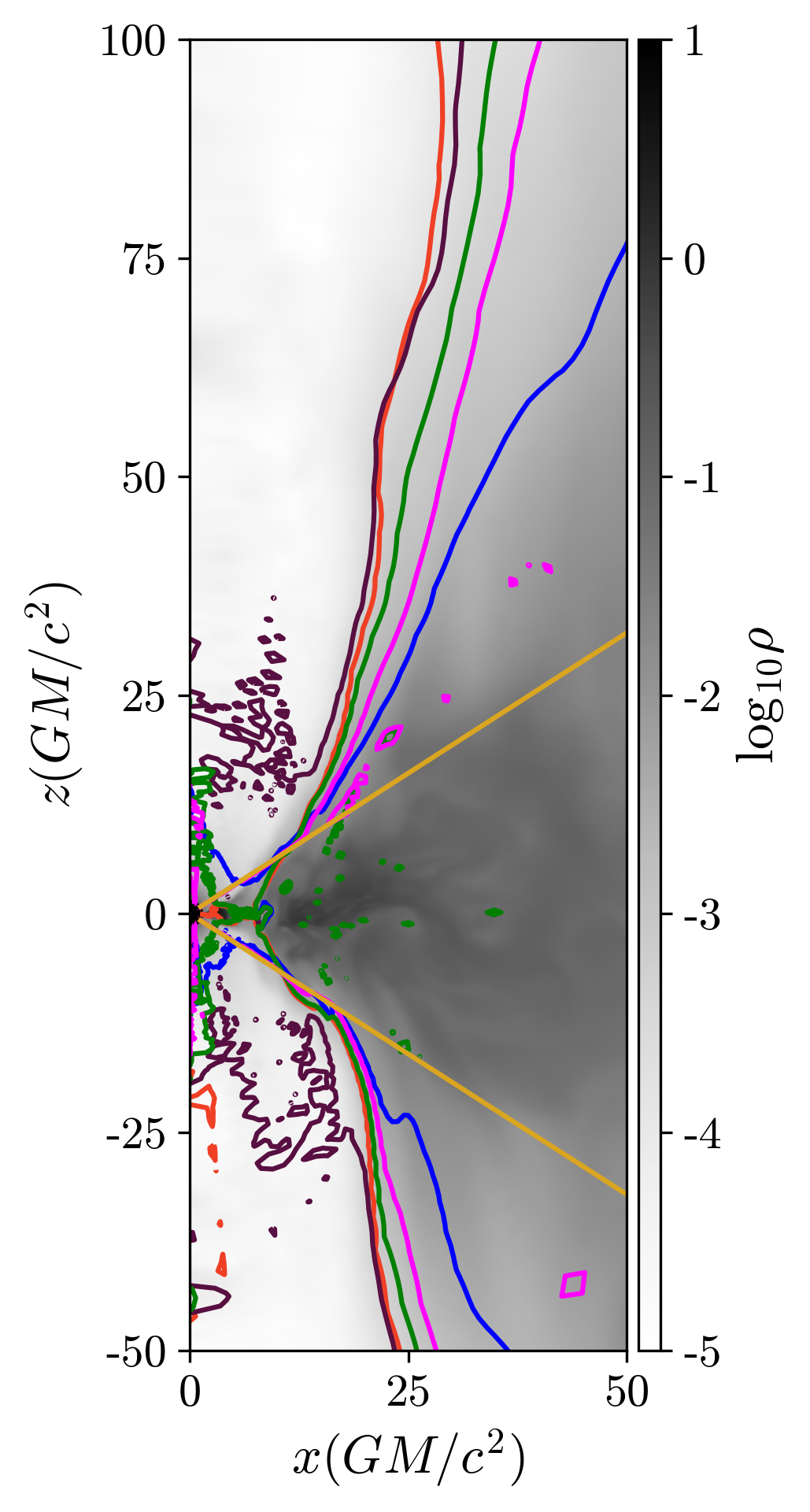}
\caption{A poloidal cut of a MAD simulation illustrating various definitions of a jet or outflow. The background color saturation represents rest-mass density in code units. The goldenrod lines mark geometric boundaries at $\theta=1$ and $\theta=\pi-1$, which are used in this work to compute the total outflow power $P_{\text{out}}$. The blue line is the $-u_{t}=1$ contour, while the magenta line represents $\beta\gamma=1$, the latter serving as the jet definition in this study. The green, plum, and red lines indicate contours where $\mu=1$, $\mathrm{Be}=1$ (fluid Bernoulli parameter), and $\sigma=1$, respectively.}
\label{fig:jet_outflow_power_definitions}
\end{figure}

The overall morphology of a typical radiatively inefficient black hole accretion system is qualitatively well understood (see Figure 1 in \citealp{porth_cc_2019}), however, there is no consensus on a quantitative definition of the jet. \cite{mckinney_measurement_2004} use the geometric Bernoulli parameter $-u_{t}$, whereas \cite{narayan_sane_2012, dexter_m87_jet_2012,moscibrodzka_jets_appearance_grmhd_2014} prefer using the fluid Bernoulli parameter to classify outflows. The MHD Bernoulli parameter, which incorporates the contribution of magnetic energy, has also been used \citep{penna_torus_2013, yuan_winds_2015}. Furthermore, \cite{narayan_sane_2012} consider the ratio of total energy flux to the rest-mass flux, defined as $\mu\equiv T^{r}_{t}/(\rho u^{r})-1$. Finally, \citetalias{M87PaperV} define jet power as the total energy flux in regions of outflow over the polar caps of the black hole where energy per unit rest-mass exceeds $2.2c^2$. In this work, we adopt a jet definition similar to that used in \citetalias{M87PaperV}.

To convey a sense for the various definitions of a jet in the literature, we plot these cuts in Figure \ref{fig:jet_outflow_power_definitions} using a snapshot from the MAD $a_{*}=+0.94$ simulation.
\section{Failure modes in the \v3 library}\label{appendix:failure_modes}

\begin{figure*}
\centering
\includegraphics[,width=\linewidth]{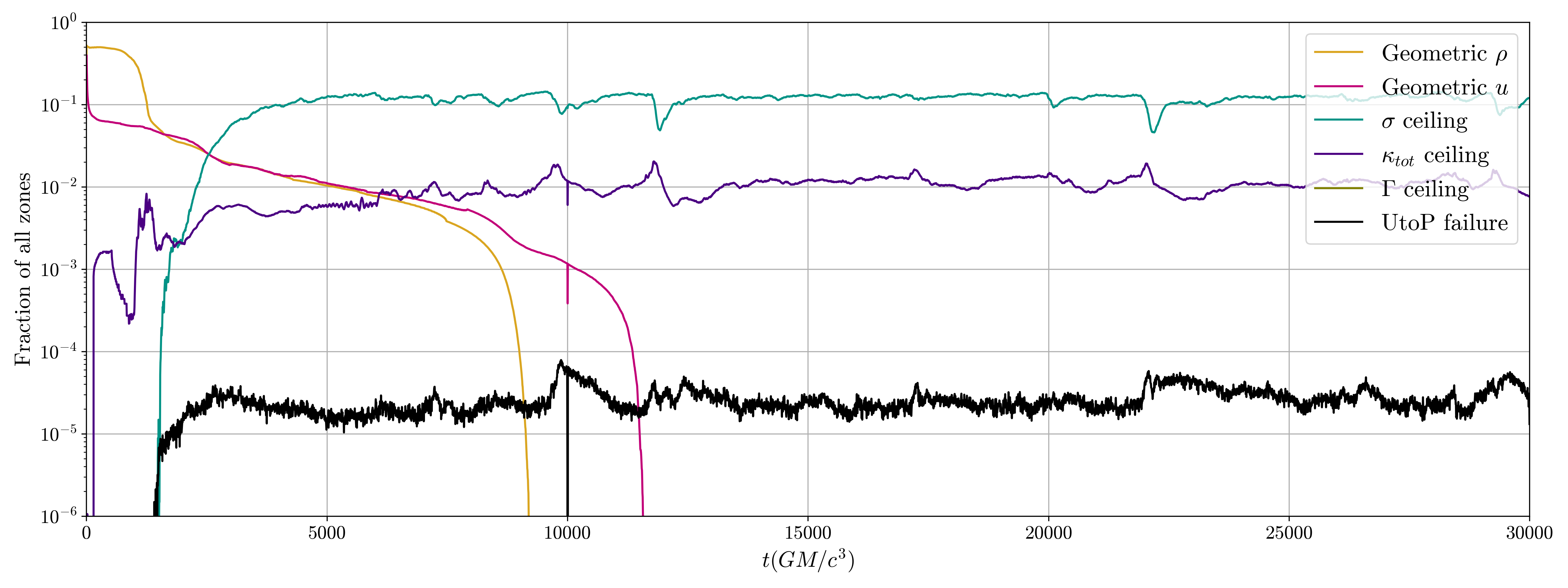}
\caption{Time series of the fraction of total zones where floors and primitive recovery failures occur for MAD $a_{*}=+0.5$. Before accretion begins, geometric floors on $\rho$ and $u$ are the dominant contribution. As the evolution progresses and the funnel region becomes magnetically dominated, $\sim$10\% of the zones have their rest-mass density set by a $\sigma$ ceiling.}
\label{fig:failure_modes_MAD}
\end{figure*}

\begin{figure*}
\centering
\includegraphics[,width=\linewidth]{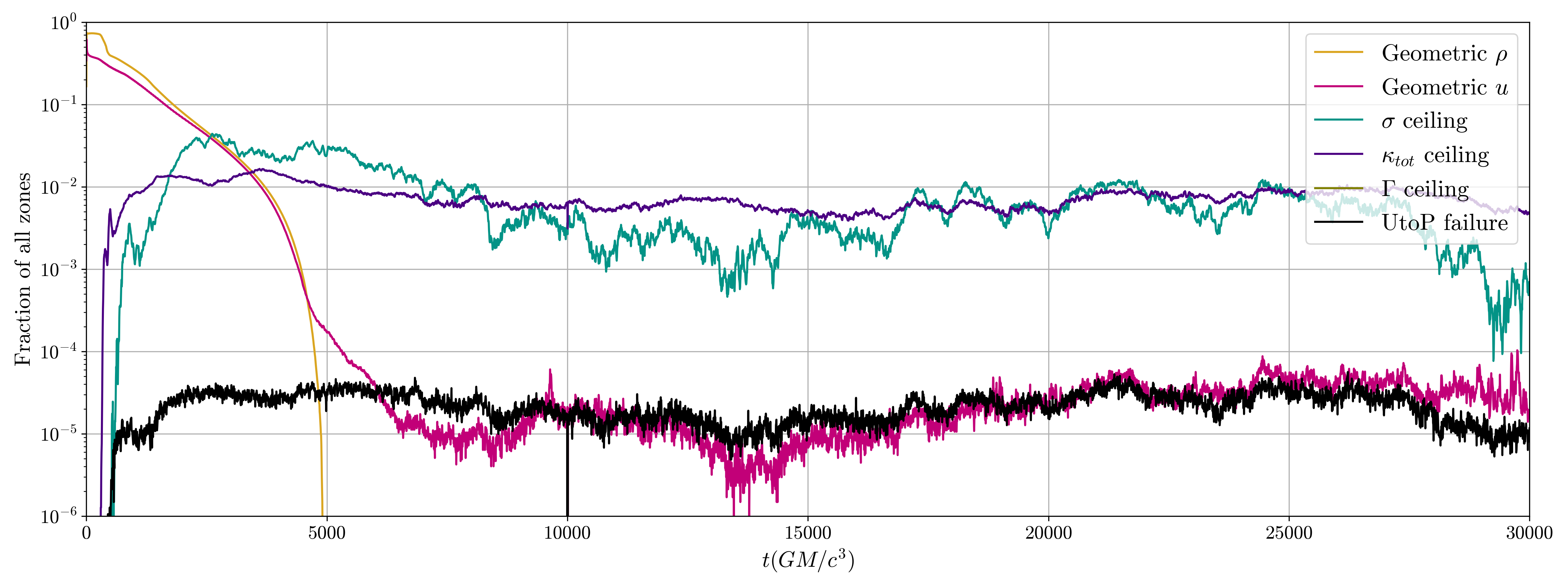}
\caption{Similar to Figure \ref{fig:failure_modes_MAD}, but for SANE $a_{*}=+0.5$. Compared to the MAD model, there are fewer flow hits, with the largest contributions coming from the entropy and $\sigma$ ceilings.}
\label{fig:failure_modes_SANE}
\end{figure*}

In highly magnetized regions ($\sigma\gg1$ or $\beta^{-1}\gg1$) the fluid rest-mass and internal energy make up only a small fraction of the total energy. The truncation error incurred when evolving the components of the stress-energy tensor in these regions can lead to significant inaccuracies in the fluid primitives. Consequently, grid-based MHD codes are prone to numerical failures during primitive variable recovery when there is an unequal contribution to the energy budget. To avoid such failures and to prevent the code from crashing, fluid quantities are reset to predefined limits. Floors on rest-mass density $\rho$ and internal energy $u$ can be applied in several ways: 
\begin{itemize}
    \item ``Geometrically'', where $\rho_{\mathrm{min}}$ and $u_{\mathrm{min}}$ are specified as a function of radius;
    \item By imposing ceilings on magnetic quantities $\sigma$ and $\beta^{-1}$, which translate into minimum values for $\rho$ and $u$,
    \item By setting limits on thermodynamic quantities such as fluid temperature $p_g/\rho$ or entropy $\kappa_{\mathrm{tot}}\equiv p_{g}/\rho^{\hat{\gamma}}$.
\end{itemize}
Typically, a combination of these floor prescriptions is employed to maintain numerical stability while minimizing their effect on fluid evolution. Additionally, we restrict the Lorentz factor $\Gamma$ to prevent superluminal speeds. In the \v3 library we consider the following floors and ceilings:
\begin{itemize}
    \item $\rho$ > $\rho_{\mathrm{min}} r^{-3/2}$ and $u$ > $u_{\mathrm{min}} r^{-5/2}$, where $\rho_{\mathrm{min}} = 10^{-5}$ and $u_{\mathrm{min}} = 10^{-7}$,
    \item $b^{2}/\rho < \sigma_{\mathrm{max}}$, where $\sigma_{\mathrm{max}}=100$,
    \item $\kappa_{\mathrm{tot}} < \kappa_{\mathrm{max}}$ where $\kappa_{\mathrm{max}}=3$,
    \item $\Gamma < \Gamma_{\mathrm{max}}$ where $\Gamma_{\mathrm{max}}=50$.
\end{itemize}

Floors can be applied in various frames, e.g., the comoving fluid frame \citep{gammie_harm_2003}, the normal observer frame \citep{mckinney_general_2012}, or the drift frame \citep{ressler_sgra_electrons_2017}. The choice of frame affects the stability of the scheme in highly magnetized regions. Fluid frame floors directly modify the rest-mass density and internal energy primitives and are typically less stable than those applied in the normal observer or drift frame. In the \v3 library, we inject material in the normal observer frame, which modifies the fluid conserved variables. This is followed by a $\boldsymbol{U}\rightarrow\boldsymbol{P}$ operation for the floored grid zones. If any zone fails during the  $\boldsymbol{U}\rightarrow\boldsymbol{P}$ operation, \kharma computes an average over adjacent zones where the primitives were successfully recovered.

\begin{figure}
\centering
\includegraphics[,width=\linewidth]{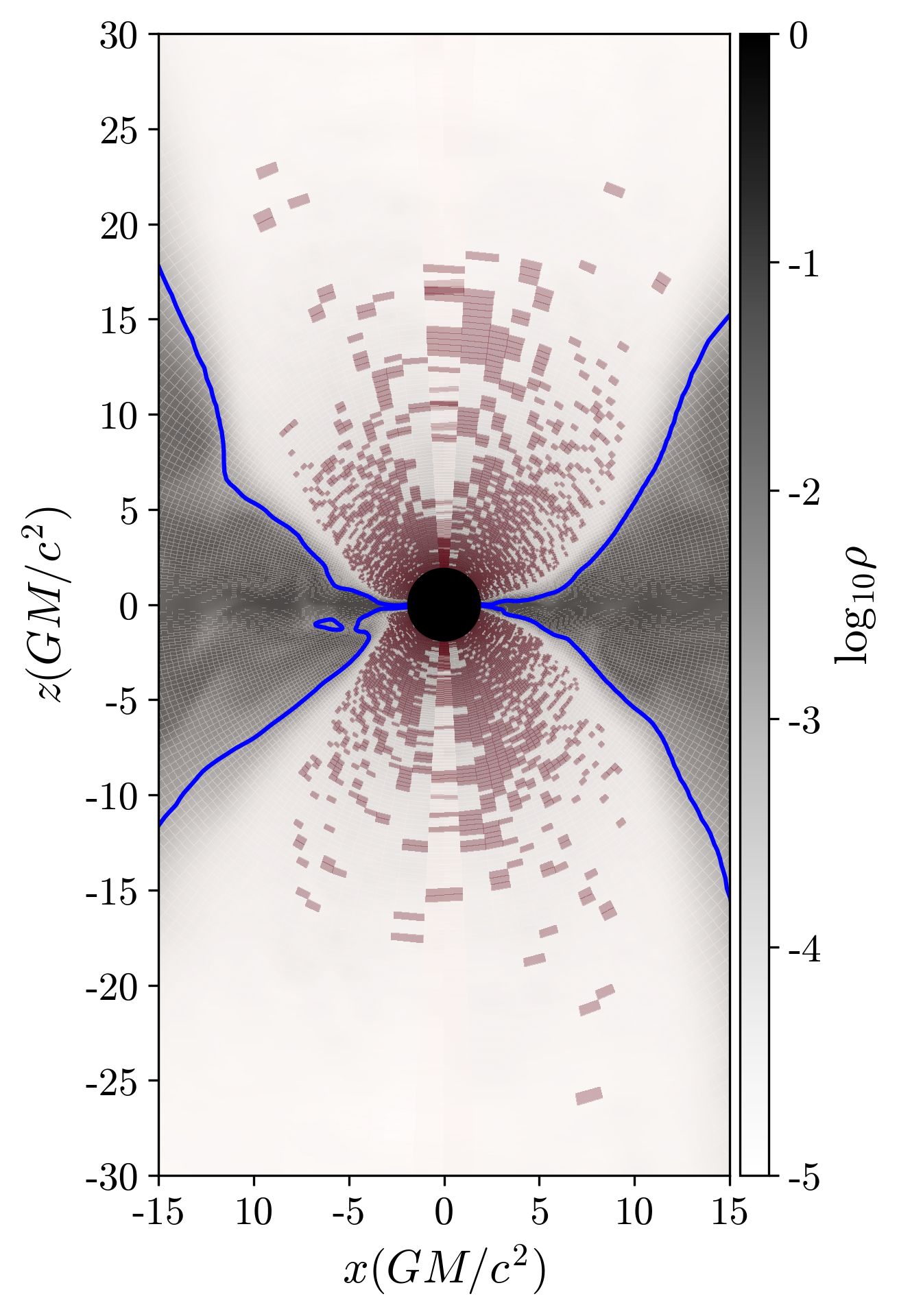}
\caption{A snapshot from the MAD $a_{*}=+0.5$ simulation. Grid zones where $\sigma$ exceeds $\sigma_{\mathrm{max}}$ are highlighted in red, while the blue contour indicates $\sigma=1$. The underlying grayscale
shading shows the logarithm of the rest-mass density in code units. The exponential radial coordinate focuses zones near the event horizon. Since $\sigma$ peaks at the base of the jet, a substantial fraction of grid zones reach the $\sigma$ ceiling in MAD simulations (see Figure \ref{fig:failure_modes_MAD}).}
\label{fig:sigma_ceiling_hits_MAD}
\end{figure}

In Figures \ref{fig:failure_modes_MAD} and \ref{fig:failure_modes_SANE} we show the time series of the fraction of total zones where the floors and ceilings are triggered for two of the models, MAD and SANE $a_{*}=+0.5$ respectively. Initially, the magnetic field is confined to the FM torus, with only geometric floors activated in the regions outside the torus. As accretion begins, jets are launched, triggering the $\sigma$ ceiling. MAD accretion flows produce powerful jets, with $b^{2}/\rho\gg1$ in the evacuated funnel region, explaining the larger fraction of zones reaching the $\sigma$ ceiling in MAD simulations (see, e.g., Figure \ref{fig:sigma_ceiling_hits_MAD}). For the chosen value of $\kappa_{\mathrm{max}}$, approximately 1\% of the zones hit the entropy ceiling, which limits the internal energy in the funnel. Additionally, the ceiling on the Lorentz factor is primarily triggered in the MAD $a_{*}=+0.94$ simulation, as high-spin MADs generate the most powerful jets (see Section \ref{sec:jet power}), accelerating particle to large velocities in the funnel. Finally, we plot the fraction of zones where $\boldsymbol{U}\rightarrow\boldsymbol{P}$ fails and find that the failure rate is at a sub-percent level.

We conclude by highlighting several stability features introduced in \kharma since the \v3 library was generated, which reduces the use of floors and makes the code more robust in highly-magnetized regions. \kharma now defaults to the primitive recovery scheme outlined in \cite{kastaun_recovery_2021}, which guarantees convergence to a valid and unique solution. This eliminates the need for a fixup routine and minimizes artificial modification to the fluid primitives. \kharma now supports ``first-order flux correction'' \citep{beckwith_mhd_methods_2011}, which attempts a speculative step using higher-order reconstruction methods and reverts to a first-order reconstruction method in zones prone to floor application. This maintains the conservative nature of the scheme at the expense of some numerical dissipation.

\bibliography{main}
\bibliographystyle{aasjournal}

\end{document}